\documentclass[trackchanges,twocolumn]{aastex701}

\usepackage[version=4]{mhchem}
\usepackage{booktabs}
\usepackage{tabularx}
\usepackage{makecell}
\usepackage{array}
\usepackage{caption}

\begin{document}

\title{Assessing the Impact of Varying HSO Cross Sections on Photochemical Models: Implications for the Spectral Characterization of Terrestrial Exoplanets}

\author[orcid=0009-0006-2271-7741]{Alexandre Branco}
\affiliation{Institute of Astrophysics and Space Sciences, Universidade do Porto, CAUP, Rua das Estrelas, 4150-762 Porto, Portugal}
\affiliation{Departamento de Física e Astronomia, Faculdade de Ciências, Universidade do Porto, Rua do Campo Alegre, 4169-007 Porto, Portugal}
\affiliation{Bard College, 30 CampusRd, Annandale-On-Hudson, NY 12504, USA}
\email[show]{alexandre.branco@astro.up.pt}

\author[orcid=0000-0002-7853-6871]{Clara Sousa-Silva}
\affiliation{Institute of Astrophysics and Space Sciences, Universidade do Porto, CAUP, Rua das Estrelas, 4150-762 Porto, Portugal}
\affiliation{Bard College, 30 CampusRd, Annandale-On-Hudson, NY 12504, USA}
\email{csousasilva@bard.edu}

\author[orcid=0009-0006-2304-3419]{Wynter Broussard}
\affiliation{Department of Earth and Planetary Sciences, University of California, Riverside, CA 92521, USA}
\email{abrou009@ucr.edu}

\author[orcid=0000-0002-2949-2163]{Edward W. Schwieterman}
\affiliation{Department of Earth and Planetary Sciences, University of California, Riverside, CA 92521, USA}
\affiliation{Blue Marble Space Institute of Science, Seattle, WA 98104, USA}
\email{eschwiet@ucr.edu}

\author[orcid=0000-0002-5147-9053]{Sukrit Ranjan}
\affiliation{University of Arizona, Lunar and Planetary Laboratory/Department of Planetary Sciences, Tucson, AZ 85721, USA}
\affiliation{Blue Marble Space Institute of Science, Seattle, WA 98104, USA}
\email{sukrit@arizona.edu}

\author{Vera Topcik}
\affiliation{Bard College, 30 CampusRd, Annandale-On-Hudson, NY 12504, USA}
\email{vt0082@bard.edu}

\author[orcid=0000-0001-7918-0355]{Olivier D. S. Demangeon}
\affiliation{Institute of Astrophysics and Space Sciences, Universidade do Porto, CAUP, Rua das Estrelas, 4150-762 Porto, Portugal}
\affiliation{Departamento de Física e Astronomia, Faculdade de Ciências, Universidade do Porto, Rua do Campo Alegre, 4169-007 Porto, Portugal}
\email{olivier.demangeon@astro.up.pt}

\author[orcid=0000-0001-6757-5763]{Pedro Machado}
\affiliation{Institute of Astrophysics and Space Sciences, Observatório Astronómico de Lisboa, Ed. Leste, Tapada da Ajuda, 1349-018 Lisboa, Portugal}
\affiliation{Faculdade de Ciências, Universidade de Lisboa, Campo Grande 016, 1749-016 Lisboa, Portugal}
\email{pmmachado@fc.ul.pt}

\begin{abstract}

Characterization of exoplanet atmospheres requires a close interplay between observations, modelling and experimental data. The accuracy of input data used in atmospheric models is essential, as it impacts our interpretation of planetary spectra with retrieval codes. Molecular absorption cross sections in the Ultraviolet-Visible range are fundamental input parameters, determining chemical kinetics, particularly on temperate terrestrial planets. However, several atmospheric species remain poorly, or even entirely uncharacterised. This is the case for HSO, a radical with unconstrained photolysis cross sections, often approximated by hydroperoxyl (HO$_2$). Sulphur chemistry can strongly influence the composition of rocky exoplanets, particularly in anoxic environments where volcanic SO$_2$ -- the main source of HSO -- is more efficiently photolysed, and sulphur aerosols like S$_8$ can form. HSO photolysis contributes to the formation of such sulphur chains, which have been proposed as indirect signatures of volcanic outgassing. Assessing the sensitivity of photochemical models to different UV–Visible cross-section prescriptions for HSO is therefore important for guiding its prioritization among poorly characterised atmospheric species. Here, we derive an updated HSO cross-section prescription from simulated HSO$_2$ data, providing a more reliable representation of HSO photolysis than HO$_2$. We compare results from these new cross sections to the default prescription for temperate terrestrial planets with Archean-like atmospheres. We find that our updated HSO cross-section prescription enhances aerosol scattering and absorption signatures in transmission, emission and reflection spectra for planets orbiting G- and K-type stars.


\end{abstract}


\section{Introduction} \label{sec:Intro}

Characterizing the atmospheres of small, rocky exoplanets with secondary atmospheres is among the most significant challenges in modern astronomy. JWST currently provides the most promising path to probing the atmospheres of planets down to the Earth-sized regime, offering an unprecedented opportunity to answer big questions in exoplanetary science, such as whether terrestrial exoplanets orbiting M-dwarf stars can sustain atmospheres at all \citep[e.g.,][]{Glidden2025, Piaulet-Ghorayeb2025}. Upcoming spectroscopic facilities including the European Extremely Large Telescope (E-ELT), the Habitable Worlds Observatory (HWO), and the Large Interferometer For Exoplanets (LIFE), are expected to enable increasingly detailed studies of terrestrial exoplanet atmospheres in the near future, offering the capability to understand exoplanet diversity and search for biosignatures on temperate, rocky worlds around Sun-like stars \citep{Quanz2022,DecadalSurveyHWO,Palle2025}.

Photochemical studies, combined with forward modelling, offer the possibility to anticipate and interpret spectroscopic observables originating from distinct terrestrial exoplanet environments. Ultimately, the robustness of photochemical models relies on accurate input data, including stellar spectra, chemical reaction rates, molecular absorption cross sections, dry and wet deposition rates, and mixing parametrizations \citep{Fortney2019}. Importantly, numerous molecular species remain poorly constrained or entirely uncharacterised in terms of their absorption cross sections \citep[e.g.,][]{sousa2019molecular}. In the complete absence of data, the inclusion of such molecules into photochemical models is often done by proxy, that is, by working with data from a different species which has presumably close absorption cross sections to the molecule of interest. Consequently, the integrity of photochemical networks that rely on such assumptions is uncertain, potentially compromising observation planning and leading to erroneous results from atmospheric retrieval codes, such as misassignments of spectral features or estimates of inaccurate atmospheric abundances \citep[e.g.,][]{Niraula2022}.

Recent studies have shown that updates to H$_2$O absorption cross sections extending into the mid-UV (MUV; 200–300 nm) can significantly impact predictions of trace-gas chemistry and observable spectral features for temperate terrestrial exoplanets with anoxic, Archean-like atmospheres \citep{Ranjan2020,Broussard2024}. In contrast, updates to CO$_2$ MUV prescriptions generally have minimal predicted observational consequences, with the exception of planets orbiting M-type stars, where the appearance of abiotic ozone features in the infrared is sensitive to the adopted cross sections \citep{Broussard2025}. Such results highlight the urgent need to prioritize specific molecules for absorption cross-section characterization via high-accuracy experimental measurements or {\it{ab initio}} calculations, both of which require significant time and resources \citep[e.g.,][]{tennyson2016exomol}.

HSO is one such species whose UV-Visible absorption cross sections remain entirely uncharacterised, with its photolysis often being modelled assuming that it can be approximated to that of hydroperoxyl (HO$_2$) \citep{PavlovKasting2002,Hu2012,Hu2013}, despite the consequences of this assumption being untested. As an atmospheric radical, HSO is known to take part in photochemical networks involving sulphur-bearing species, such as SO$_2$ and H$_2$S -- common volcanic gases on Earth which may occur in even greater concentrations on some rocky exoplanets \citep{Patel2024,Bello-Arufe2025}. While serving as the main driver of HSO formation, the photochemical processing of SO$_2$ plays a key role in shaping the spectra of distinct terrestrial bodies in the Solar System, including Venus and Io, and is also known to have significantly influenced Earth’s atmospheric composition during the Archean eon, between 4--2.5 Gyr ago \citep[e.g.,][]{CatlingZahnle2020,Hu2013}.

In anoxic environments, similar to that of the Archean Earth, stellar MUV photons penetrate the troposphere, driving the photolysis of volcanic SO$_2$. This contrasts with O$_2$-rich atmospheres, where most of these photons are shielded by stratospheric O$_2$ and O$_3$. Under the reducing conditions of an Archean-like atmosphere, species such as S, S$_2$, or S$_3$ can persist and polymerize into water-insoluble S$_8$ aerosols. Except for S$_8$ and sulphuric acid (H$_2$SO$_4$) aerosols, most sulphur-bearing gases are moderately soluble and susceptible to UV photolysis, thus remaining short-lived. As a result, atmospheric sulphur is sequestered into stable H$_2$SO$_4$ aerosols by oxidation, or S$_8$ aerosols by reduction.

There are two primary channels for the photolysis of HSO:
\begin{equation}
\label{react:HSO_phot_1}
\ce{HSO + $h\nu$ -> HS(X^{2}\Pi) + O(^3P_g)}
\end{equation}
\vspace{-13pt}
\begin{equation}
\label{react:HSO_phot_2}
\ce{HSO + $h\nu$ -> SO(X^3\Sigma^-) + H(^2S_g)}
\end{equation}
HSO photolysis, following Equation \ref{react:HSO_phot_1}, is a particularly important intermediate step in the formation of S$_8$, as it provides a secondary source of HS radicals, primarily derived from the photolysis of H$_2$S \citep{Hu2013}. HS radicals can react with H or with another HS to yield atomic S, or combine with S to form S$_2$, which serve as building blocks for longer sulphur chains.

Inferring volcanic sulphur emissions on terrestrial exoplanets with anoxic atmospheres from the direct detection of chemically short-lived SO$_2$ or H$_2$S is expected to be challenging. \cite{Hu2013} estimated that sustaining detectable levels of SO$_2$ and H$_2$S in planetary spectra would require surface emission fluxes 1000 and 3000 times greater, respectively, than present-day Earth’s global volcanic sulphur flux. However, even if atmospheric concentrations of SO$_2$ and H$_2$S remain below detectable levels, photochemical conversion to stable S$_8$ aerosols could provide an indirect signature of ongoing volcanic outgassing, with HSO absorption cross sections potentially being consequential for their detectability.

In this work, we aim to demonstrate the sensitivity of photochemical models of anoxic terrestrial atmospheres to updated HSO cross sections, assessing the need for detailed characterization of the HSO UV–Visible spectrum. Based on structural chemistry considerations, we identify a new proxy molecule and propose an updated HSO cross-section prescription for more accurate modelling of HSO photolysis than that achievable with HO$_2$ data. We test the sensitivity of these inputs under a range of SO$_2$ surface fluxes for FGK-type dwarf host stars. In Section \ref{sec:Methods}, we describe the methods used to construct the updated HSO cross sections. Additionally, we describe the photochemical and spectral models, as well as the planetary scenario used to test the sensitivity of these inputs. In Section \ref{sec:Results}, we report our results, including impacts on trace gas species and simulated emission, reflection, and transmission spectra. We discuss the implications of our results in Section \ref{sec:Discussion} and conclude in Section \ref{sec:Conclusions}. 

\section{Methods} \label{sec:Methods}
\subsection{Towards New Cross-Section Data}

To perform HSO cross-section sensitivity tests, we investigated alternative proxy molecules whose cross sections could enable more accurate modelling of HSO photolysis than what is possible with standard HO$_2$ data. We note that no substitute molecule can fully replicate the spectral behaviour of HSO. However, identifying proxy data that more faithfully captures the UV-Visible absorption characteristics of HSO reduces the modelling uncertainty inherent to the use of HO$_2$ data, ultimately enabling tests of how strongly model outcomes depend on the assumed cross sections.

\setlength{\tabcolsep}{14pt}
\begin{deluxetable}{lccccccc}
\tabletypesize{\footnotesize}
\tablewidth{\textwidth} 
\tablecaption{Fundamental vibrational frequencies of HO$_2$, HSO, and proxy molecules considered in this study. \label{table:Proxies_Vib_Freq}}
\tablehead{
\colhead{\textbf{Vib. Freq. [cm$^{-1}$]}} & 
\colhead{$\nu_1$} & 
\colhead{$\nu_2$} & 
\colhead{$\nu_3$} & 
\colhead{$\nu_4$} & 
\colhead{$\nu_5$} & 
\colhead{$\nu_6$} & 
\colhead{Ref.}
}
\startdata
\textbf{HO$_2$} & 3436.2 & 1391.8 & 1097.6 & -- & -- & -- & J1998 \\
\textbf{HSO} & 2335.1 & 1076.8 & 1002.3 & -- & -- & -- & D2009 \\
\textbf{HSO$_2$} & 2168.0 & 1299.3 & 1084.0 & 958.9 & 799.7 & 464.8 & F2019 \\
\textbf{HOSO} & 3567.3 & 1172.2 & 1034.1 & 781.0 & 395.6 & 31.0 & L2021 \\
\textbf{SO$_2$} & 1151.4 & 517.7  & 1361.8 & -- & -- & -- & P1982 \\
\textbf{SO} & 1138.0 & -- & -- & -- & -- & -- & I2007 \\
\enddata
\tablecomments{References: J1998: \cite{Jacox1998}; D2009: \cite{Denis2009}; F2019: \cite{Fortenberry2021}; L2021: \cite{Lu2021}; P1982: \cite{Person1982}; I2007: \cite{Irikura2007}.}
\end{deluxetable}

\subsubsection{Functional Groups}

Organic chemists have long recorded infrared molecular spectra and assigned specific features to their corresponding functional groups, that is, groups of atoms and bonds whose collective motion determines a molecule’s spectral properties \citep{sousa2019molecular}. Here, we adopt this approach to predict the spectral behaviour of HSO in the UV-Visible range, by comparing its fundamental vibrational modes to those of molecules with shared functional groups. This analysis relies on the assumption that similarities in infrared vibrational features, arising from common functional groups, may in principle translate to comparable spectral activity in the UV-Visible. However, we must point out that this correspondence is not guaranteed, and potential deviations should be expected.

To the best of our knowledge, only one fundamental vibrational frequency of HSO has been experimentally observed, corresponding to its SO stretching mode \citep{Schurath1977}, while its remaining vibrational frequencies have been theoretically determined \citep{Denis2009}. Despite the limited spectral data available for HSO, if we are to make the argument of using functional groups to identify a proxy molecule, not only does the SO functional group emerge as a natural starting point, but it also highlights HO$_2$ as an illogical choice. Table \ref{table:Proxies_Vib_Freq} compares the fundamental vibrational frequencies of HO$_2$, HSO and potential proxy molecules for the latter, including HSO$_2$, HOSO, SO$_2$ and SO. Importantly, any suitable proxy molecule should be an SO-bearing species and meet the basic requirement of having available cross-section data in the UV-Visible range.

By comparing the fundamental vibrational frequencies of HSO with those of potential proxy molecules, we observe that both HSO$_2$ and its isomer, HOSO, exhibit the most similar fundamental frequencies to those of HSO. In contrast, SO$_2$ and SO show significant differences, which rules them out as potential proxies. Considering that the differences in vibrational frequencies between HSO$_2$, HOSO, and HSO vary across different vibrational modes -- with HSO$_2$ being significantly closer to HSO in the first mode, but HOSO approaching HSO more closely in the second and third modes -- we further distinguish between the suitability of HSO$_2$ and HOSO as proxy molecules by considering molecular symmetry arguments.

\subsubsection{Molecular Symmetry}

The UV-Visible absorption spectrum of most molecules is dominated by electronic transitions. Molecular symmetry plays a crucial role in this process, as it imposes important constraints on molecular orbitals and governs the allowed transitions. Consequently, when looking for a suitable proxy for HSO, it is important to choose a molecule whose symmetry allows for similar electronic transitions. Whether a transition is allowed can be determined using detailed symmetry arguments formulated within the framework of group theory. While much of group theory formalizes intuitive ideas about the symmetries of objects, its rules can be applied in a systematic way to discuss molecular properties \citep{AtkinsPaula2010}.

Molecules with similar properties are typically grouped together by identifying their symmetry elements \citep{AtkinsPaula2010}. A symmetry element takes the form of a point, line or plane with respect to which it is possible to perform an action over a molecule that leaves it looking the same after it has been carried out. The classification of molecules according to symmetry elements corresponding to operations that leave at least one common point unchanged (i.e., rotations, translations, or inversions) gives rise to point groups \citep{AtkinsPaula2010}.

The HSO molecule belongs to the $C_{s}$ point group \citep{Li2024}, meaning it has a plane of reflection through which an identical copy of the original molecule is generated. For a detailed explanation of point group notation we refer to \cite{AtkinsPaula2010}. Importantly, while HSO$_2$ is part of the $C_{s}$ point group \citep{Fortenberry2021}, its isomer HOSO lacks any symmetry elements, meaning that no operation, besides the identity operation, preserves a single point after being carried out. Keeping this in mind, we identify HSO$_2$ as the most suitable proxy molecule to assess the impact of HSO cross-section data on exoplanet characterization.

\subsubsection{Cross-Section Prescriptions}

\begin{figure*}[ht!]
\centering
\includegraphics[width=0.82\textwidth]{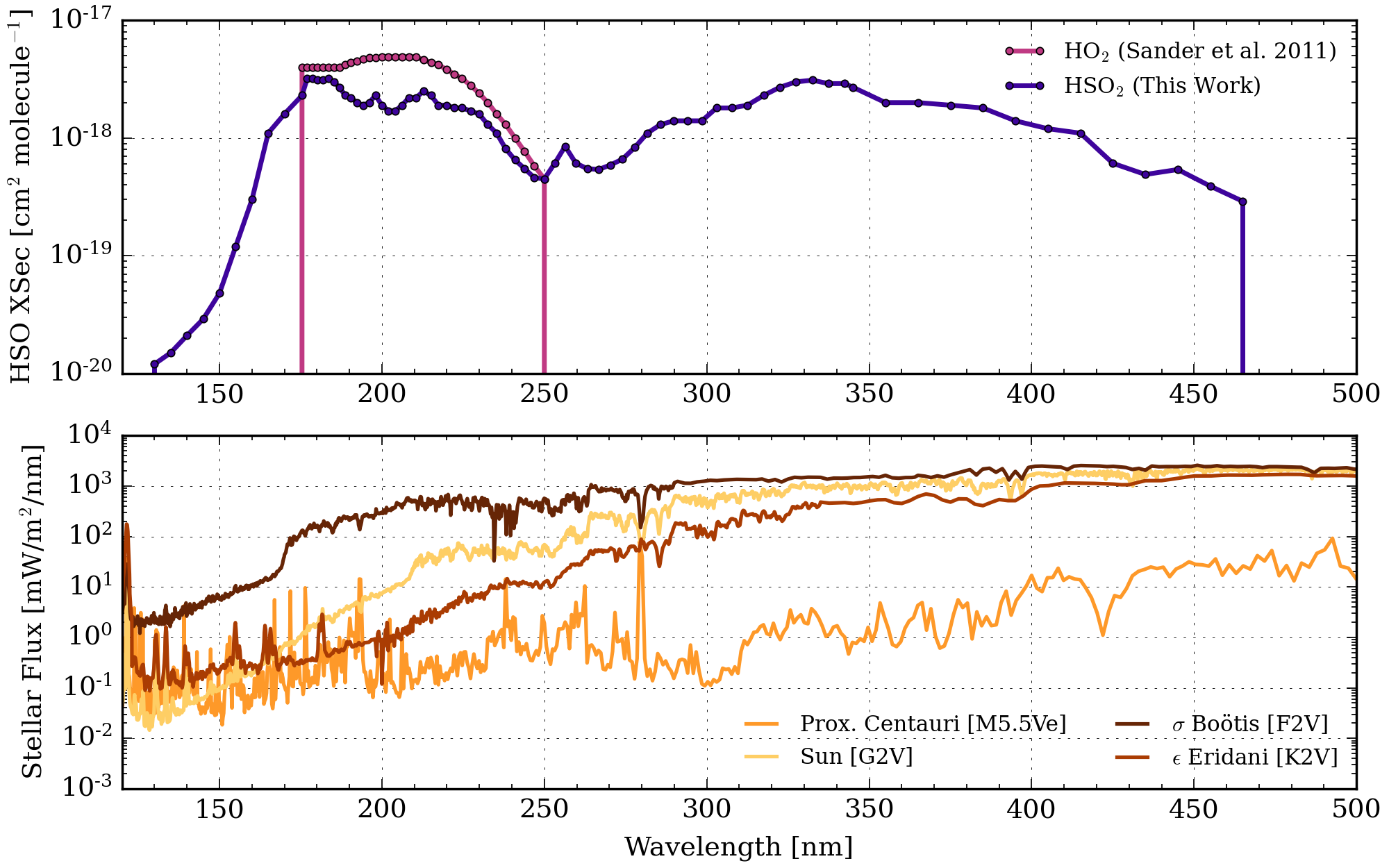}
\caption{Top: HSO cross-section prescriptions from 120 to 500 nm. Bottom: Spectral energy distributions of the FGK-type stars used in the sensitivity tests, together with an M-type stellar spectrum.\label{fig:stellar_flux_hso_xsec}}
\end{figure*}

We then conduct HSO cross-section sensitivity tests using the two prescriptions, shown in Figure \ref{fig:stellar_flux_hso_xsec}. This includes the standard HO$_2$ cross sections from \cite{Sander2011}, which serve as the default in the Atmos photochemical model (see Section \ref{sec: Atmos}), and an updated HSO cross-section prescription derived from the simulated UV-Visible absorption spectrum of HSO$_2$ from \cite{Lu2021}.

To derive our new prescription we recall that the HSO photolysis channel in Equation (\ref{react:HSO_phot_1}) has a quantum limit at 302.4 nm, while the channel in Equation (\ref{react:HSO_phot_2}) has a quantum limit at 466.1 nm, representing the energies needed to break the SH--O and SO--H bonds in the ground state, respectively. The pathway in Equation (\ref{react:HSO_phot_2}) thus requires the minimum energy to trigger HSO photolysis. Beyond this quantum limit, only forbidden transitions can occur. Importantly, recent work by \cite{Broussard2024} has shown that the accumulation of these forbidden transitions can still add to the opacity of H$_2$O absorption cross sections, significantly impacting photochemical networks, and leading to observable changes in simulated planetary spectra. The observational consequences of such transitions can vary depending on the molecule, atmospheric composition, and host star's spectral type, and may in some cases be minimal \citep{Broussard2025}. In the current work, we do not account for this cumulative opacity effect beyond 466.1 nm (e.g., through the use of extrapolations to predict this data), as we did not want to presume that the near-dissociation behaviour of HSO could be reliably estimated from our chosen proxy.

Additionally, HSO photolysis in the current Atmos photochemical network is only implemented through the channel in Equation (\ref{react:HSO_phot_1}), which directly feeds into the sulphur polymerisation pathways leading to S$_8$ formation in Archean-like atmospheres (see Section \ref{sec:Intro}). In this context, we adopt a constant quantum yield of 1 for this channel across the full wavelength range of the updated prescription. Although the previously mentioned quantum limits suggest that a wavelength-dependent branching between the two channels would be physically motivated, implementing a second photolysis pathway in the network would add an additional layer of complexity relative to previous HSO photolysis prescriptions, which similarly assume a quantum yield of 1 for the HS + O channel across the wavelength range where the default HSO proxy absorbs. Our approach isolates the effect of the revised cross sections themselves, enabling direct comparison with the default prescription without the influence of an additional photolysis channel.

We propose a new HSO cross-section prescription by truncating UV-Visible absorption cross sections for HSO$_2$ \citep{Lu2021} at 466.1 nm, slightly refining the proxy data to better reflect the photochemical behaviour expected for HSO. Notably, our updated prescription extends further beyond the HO$_2$ cut-off and into the longer wavelengths -- spectroscopic behaviour which we expect to be characteristic of SO-bearing species such as HSO.

For the purpose of these sensitivity tests, we used the updated H$_2$O cross-section prescription from \cite{Ranjan2020} and the CO$_2$ cross sections from \cite{KastingWalker1981}. Our updated HSO cross-section prescription is available at \dataset[doi:10.5281/zenodo.20209640]{https://doi.org/10.5281/zenodo.20209640} and \url{https://github.com/alexoworlds/branco_2026_hso}.

\subsection{Photochemical Model Description} \label{sec: Atmos}

This work makes use of the photochemical component of \texttt{Atmos}, a one-dimensional coupled photochemical-climate model, frequently used to model terrestrial exoplanet atmospheres \citep{Kasting1979, Felton2022, Schwieterman2022, Broussard2024, Broussard2025}.

The photochemical module of \texttt{Atmos} takes in prescribed planetary and stellar parameters, including the planet’s atmospheric pressure–temperature profile, the stellar flux at the relevant planet–star distance, and gas species boundary conditions.

\texttt{Atmos} approximates the atmosphere to a set of plane-parallel, vertically stratified layers, where vertical mixing is parametrized. Our model atmosphere is divided into 200 plane-parallel layers from the surface to 100 km in altitude, with a layer spacing of 0.5 km. The steady-state volume mixing ratio of each species in the atmosphere is determined by solving the coupled one-dimensional continuity–transport equation, where vertical transport is approximated as a combination of molecular and eddy diffusion. Both flux and mass continuity equations are solved at each atmospheric layer using the inverse-Euler method, and once converged, the code returns volume mixing ratios for relevant gases throughout the atmosphere.

In this work, we adopted the Archean + haze atmospheric template from \cite{Arney2016}, applying the boundary conditions listed in Table \ref{table:boundary_conditions} in Appendix \ref{appendixA}. The adopted pressure–temperature structure sets the surface pressure at 1 bar and the surface temperature at 275 K. The temperature decreases linearly up to the tropopause, at 11 km, above which the atmosphere is isothermal at 180 K. We note that the tropospheric water vapour profile is calculated assuming a surface relative humidity of 70\% \citep{ManabeWetherald1967}, thus different surface temperatures can lead to distinct tropospheric water content. We therefore use the same surface temperature for all planets regardless of the host star's spectral type, to facilitate direct comparisons between the host stars and the HSO cross-section prescriptions. Our pressure-temperature profile corresponds to the cool, more water-poor troposphere scenario from \cite{Broussard2024}, where reduced water content limits OH production and its effect as a sink for HSO. This choice allows differences driven by HSO cross sections to be more apparent and less influenced by variations in OH generated through water photolysis.

\subsection{Spectral Model Description}

To simulate the spectra of the modelled atmospheres we used the forward modelling component of the open source \texttt{POSEIDON} code \citep{MacDonald2017,MacDonald2023}. \texttt{POSEIDON} takes the pressure-temperature profile and altitude-dependent volume mixing ratios predicted with \texttt{Atmos} to calculate synthetic spectra for various exoplanet observing geometries, including transmission, emission, and reflection spectra. The code carries out radiative transfer calculations at each atmospheric layer, incorporating wavelength-dependent opacities for both gas-phase and aerosol species. \texttt{POSEIDON} includes an extensive opacity database with line-by-line molecular absorption cross sections and Mie scattering properties for atmospheric aerosols \citep{Li2015,Polyansky2018,Yurchenko2020,Yurchenko2024,Mullens2024}. For S$_8$ aerosols, opacities are computed using Mie theory assuming orthorhombic solid sulphur particles, with refractive indices taken from \cite{Fuller1998} as compiled in \cite{Palik1998_Handbook}. Importantly, for terrestrial planets, \texttt{POSEIDON} provides a dedicated ``temperate'' opacity database which has been optimised for improved accuracy in modelling spectra of cooler atmospheres (T $\lesssim$ 400 K).

\subsection{Stellar Spectra}

\setlength{\tabcolsep}{5pt} 
\renewcommand{\arraystretch}{1.35} 
\setlength{\extrarowheight}{4pt} 

\begin{deluxetable*}{lccccc}
    \tabletypesize{\footnotesize} 
    \tablewidth{\textwidth} 
    \tablecaption{Stellar parameters of the host stars used in the sensitivity tests.\label{table:Stellar_Properties}} 
    \tablehead{
        \colhead{\shortstack[c]{\vspace{4pt}\textbf{Star}\vspace{1pt}}} & 
        \colhead{\shortstack[c]{Spectral \\ Type}} & 
        \colhead{\shortstack[c]{$T_{\mathrm{eff}}$ \\ {[K]}}} & 
        \colhead{\shortstack[c]{Luminosity \\ {[$L_{\odot}$]}}} & 
        \colhead{\shortstack[c]{Radius \\ {[$R_{\odot}$]}}} & 
        \colhead{\shortstack[c]{Distance \\ {[pc]}}} 
    } 
    \startdata 
        \textbf{$\sigma$ Boötis} & F2V & 6540 & 3.15 & 1.43 & 15.8 \\ 
        \textbf{Sun} & G2V & 5778 & 1.00 & 1.00 & -- \\
        \textbf{$\epsilon$ Eridani} & K2V & 5039 & 0.32 & 0.735 & 3.2\\[2pt]
    \enddata 
\end{deluxetable*}

Atmospheric photochemical reactions driven by stellar flux are dependent on the host star’s spectral energy distribution. In this work, we conduct sensitivity tests for FGK-type stars, particularly the F-type star $\sigma$ Boötis \citep{Segura2003}, the Sun \citep{Thuillier2004}, and the K-type star $\epsilon$ Eridani \citep{Segura2003}. These represent bright host-star scenarios in the UV-Visible range, under which photolysis is not expected to be \textit{a priori} photon-limited, and thus facilitate sensitivity tests to variations in HSO cross-section prescriptions. Table \ref{table:Stellar_Properties} lists the stellar parameters for these host stars, and Figure \ref{fig:stellar_flux_hso_xsec} shows their spectral energy distribution in the UV-Visible wavelength range.

\subsection{Planetary Scenario}

In this work, we model N$_2$--CO$_2$--H$_2$O atmospheres, in line with
typical assumptions for major atmospheric species on habitable worlds and within the plausible bounds of the atmospheric composition of the Archean Earth \citep{Arney2016, CatlingZahnle2020}. Full atmospheric boundary conditions, including surface fluxes, surface volume mixing ratios, and dry deposition velocities (if prescribed) can be found in Table \ref{table:boundary_conditions} in Appendix \ref{appendixA}.

To perform SO$_2$ surface flux sensitivity tests, we vary the SO$_2$ flux from 3 $\times 10^9$ molecules cm$^{-2}$s$^{-1}$, which is around Earth’s global volcanic sulphur flux \citep{Seinfeld&Pandis2006}, up to 3 $\times 10^{12}$ molecules cm$^{-2}$s$^{-1}$, representing an enhancement by three orders of magnitude consistent with the minimum sulphur emission rates required for detectable SO$_2$ features, as discussed by \citet{Hu2013}.

\section{Results} \label{sec:Results}
\subsection{HSO Photolysis Sensitivity to SO$_2$ Surface Flux}

\begin{figure}[ht!]
\centering
\includegraphics[width=0.47\textwidth]{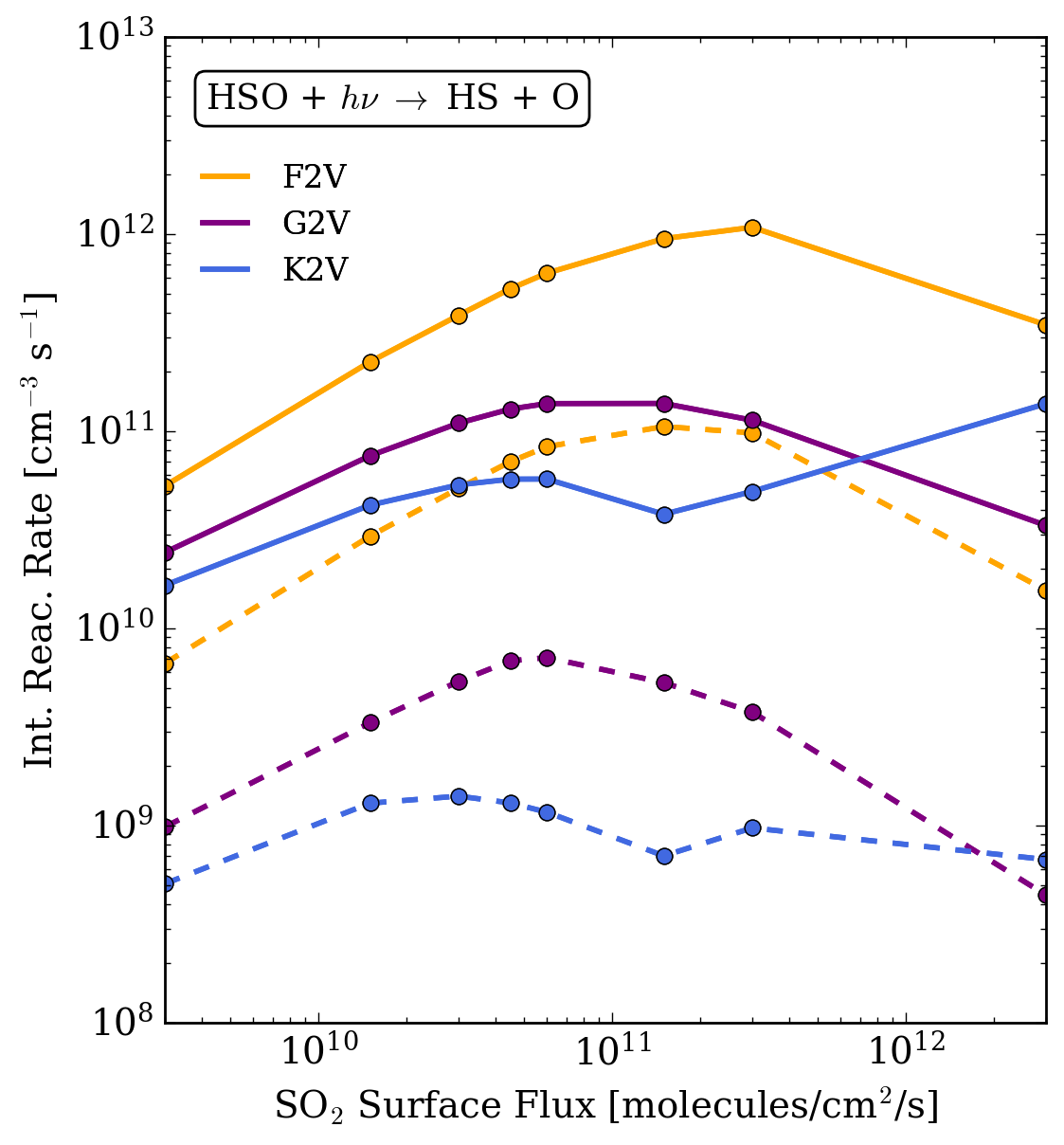}
\caption{Integrated HSO photolysis reaction rates as a function of SO$_2$ surface fluxes, ranging from $\times$1 to $\times$10$^3$ Earth’s global volcanic sulphur flux, for an anoxic planet orbiting $\sigma$ Böotis (F2V, in yellow), the Sun (G2V, in purple) and $\epsilon$ Eridani (K2V, in blue). Solid lines with circular markers show rates modelled using the updated HSO cross sections, while dashed lines with circular markers use HO$_2$ data.\label{fig:Int_HSO_PhotRate_FG}}
\end{figure}

Figure \ref{fig:Int_HSO_PhotRate_FG} shows the sensitivity of the integrated HSO photolysis reaction rates to different SO$_2$ surface fluxes for different host star types, comparing both HSO cross-section prescriptions. Additionally, we assessed the impact of scaling the updated prescription by factors of $10^3$ and $10^{-3}$, as shown in Appendix \ref{appendixB}.

Overall, the updated cross-section data yields higher HSO photolysis rates than those obtained with the previous prescription. Whereas HO$_2$ data was restricted to 175.4–250 nm, the updated cross sections span a significantly broader wavelength range, extending from 130 nm up to 465 nm. This expanded wavelength range enhances absorption both at shorter UV wavelengths and further within the visible range, ultimately increasing the integrated HSO photolysis rates.

Furthermore, we generally identify a systematic decrease in HSO photolysis rates from F- to G- to K-type stars. For some SO$_2$ surface fluxes (e.g., 3 $\times 10^{11}$ molecules cm$^{-2}$ s$^{-1}$, that is $\times$100 Earth’s global volcanic sulphur flux), the photolysis rates for G- and K-type hosts are nearly or more than an order of magnitude lower than those obtained for an F-type star across all cross-section prescriptions. This trend is expected given the progressively weaker UV–Visible flux from F- to G- to K-type stars, with a more limited supply of UV–Visible photons constraining HSO photolysis.

Additionally, we find that for the F-type host, the integrated HSO photolysis rate reaches its maximum at a surface SO$_2$ emission of $\sim$100 times Earth’s global volcanic sulphur flux. Whereas, for the G-type star, the peak occurs at $\sim$20 times Earth’s flux, SO$_2$ = 6 $\times 10^{10}$ molecules cm$^{-2}$ s$^{-1}$. This shift reflects the different UV–Visible fluxes of the two stellar types. The stronger UV–Visible flux from an F-type star can efficiently photolyse SO$_2$, form HSO, and subsequently photolyse HSO, even at high SO$_2$ surface emissions. As a result, a larger SO$_2$ flux is required before the atmosphere enters a regime where accumulated SO$_2$ absorbs enough UV–Visible radiation to shield the lower, HSO-rich layers and limit HSO photolysis. In contrast, the lower UV–Visible flux from a G-type star causes this photon-limited regime to be reached at lower SO$_2$ fluxes, shifting the peak in HSO photolysis to fluxes of $\sim$20 times Earth’s value. Past these peak fluxes, the observed decrease in the integrated HSO photolysis rate is a consequence of the system becoming fully photon-limited, as the increasingly abundant SO$_2$ absorbs a growing fraction of the incoming UV–Visible radiation.

The K-type host scenario, however, does not exhibit this clear peak followed by a decline in HSO photolysis. Instead, we observe a maximum at $\sim$15 times Earth’s flux, SO$_2$ = 45 $\times 10^{9}$ molecules cm$^{-2}$ s$^{-1}$ followed by a decline and a subsequent upturn at high SO$_2$ surface fluxes ($\sim$10$^2$–10$^3$ times Earth’s value), particularly when adopting the updated HSO cross sections. In comparison with the F- and G-type scenarios, the spectral energy distribution of the K-type star provides a weaker flux at $\lesssim$280 nm, where SO$_2$ absorbs most efficiently. As a result, HSO photolysis for a planet orbiting a K-type host is driven to a greater degree by the broader absorption range towards longer wavelengths, where SO$_2$ opacity is lower. For this stellar type, as SO$_2$ builds up and increasingly attenuates shorter wavelength radiation, it does not shield as efficiently the dominant photolysis pathway for HSO, unlike for F- and G-type stars where shorter wavelength photons carry a much larger fraction of the photolysis rate integral. The combination of a photolysis pathway that remains largely unshielded by SO$_2$, and a growing HSO reservoir derived from higher SO$_2$ surface fluxes, produces the observed upturn in the integrated photolysis rate (see Figure \ref{fig:Int_HSO_PhotRate_FG}). This behaviour is fairly less pronounced when using HO$_2$ cross-section data,  which do not extend into this SO$_2$-transparent window, supporting that the access to longer-wavelength absorption is the primary driver of the distinct K-type behaviour.

More generally, SO$_2$ flux levels where we see a maximum (local or absolute) in integrated photolysis rate correspond to planetary scenarios where HSO photolysis plays a dominant role in the chemical balance of the atmosphere, making these conditions particularly sensitive to the choice of HSO cross-section prescription and resulting in the largest potential impact on observable spectra. For this reason, we examine how the abundance profiles of trace gas species respond to different HSO cross sections in these scenarios, where the prescription's choice is expected to be most consequential.

\subsection{Relationships between HSO Cross Sections and Trace Species}

\begin{figure*}[t]
    \centering
    \begin{minipage}{0.49\textwidth}
        \centering
        \includegraphics[width=\linewidth]{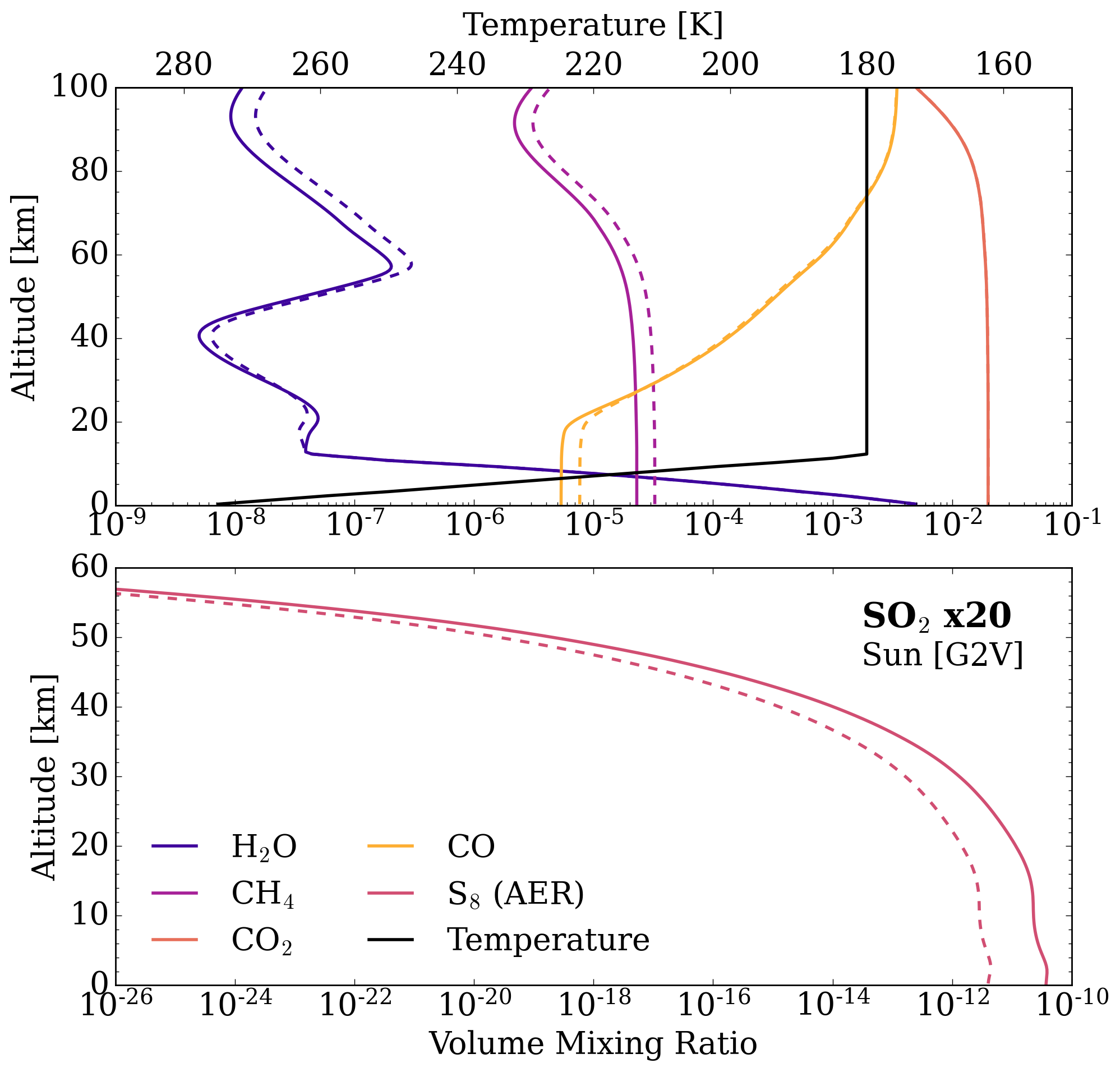}
    \end{minipage}
    \hfill
    \begin{minipage}{0.49\textwidth}
        \centering
        \includegraphics[width=\linewidth]{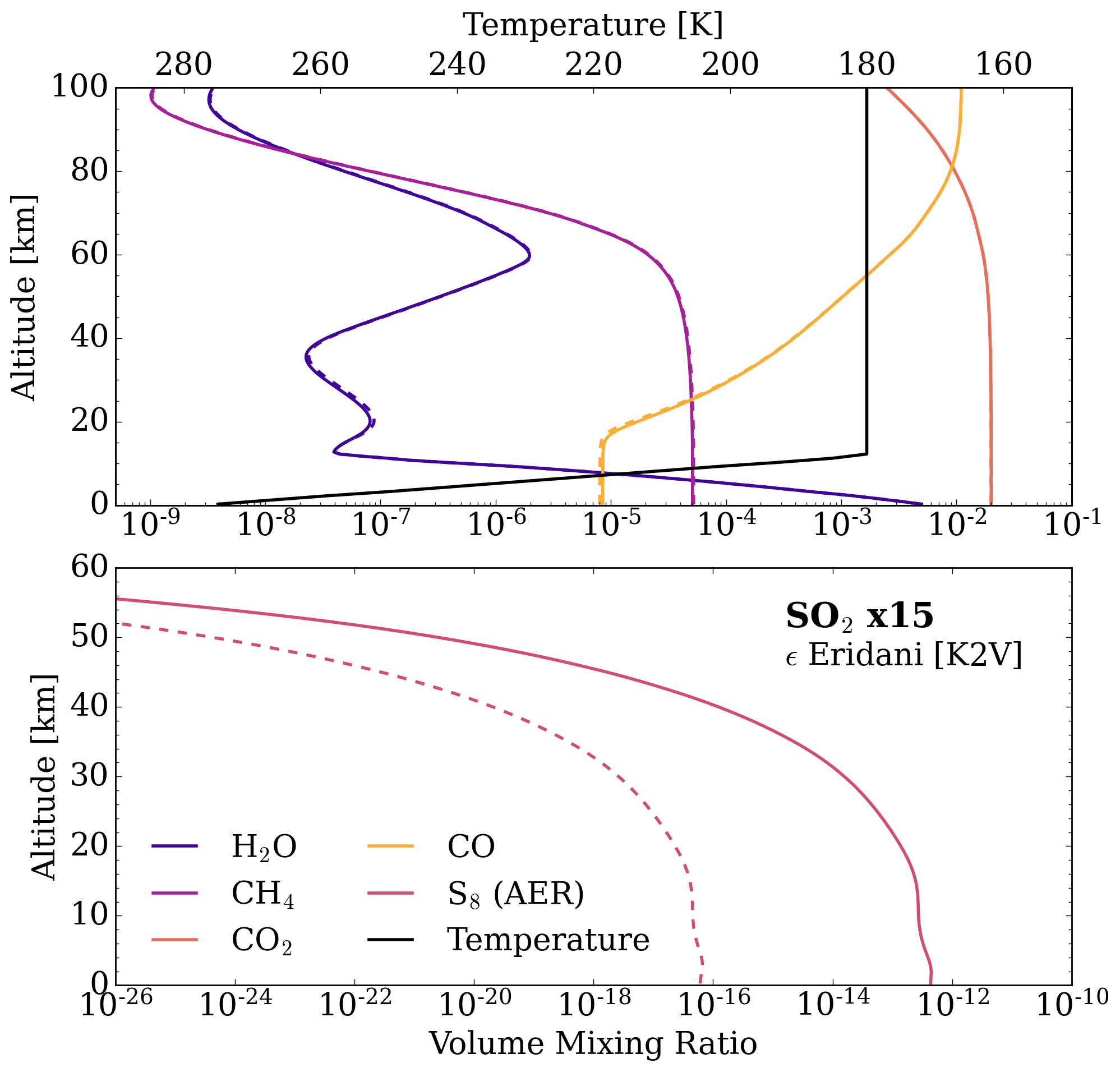}
    \end{minipage}
    \caption{\label{fig:Profiles_G_x20_F_x50} Left: Altitude-dependent volume mixing ratios of key species in an anoxic, Archean Earth-like atmosphere, orbiting the Sun, with an SO$_2$ surface flux of 6 $\times 10^{10}$ molecules cm$^{-2}$ s$^{-1}$ ($\times$20 Earth’s global volcanic sulphur flux). Solid lines show the altitude profiles modelled using the updated HSO cross sections presented in this work, while dashed lines show the profiles modelled using the default HSO prescription from \cite{Sander2011}. The abundance profile for S$_8$ aerosols (AER) up to 60 km is shown separately in the bottom panel. Right: Same as the plot on the left for a planet orbiting $\epsilon$ Eridani, with an SO$_2$ surface flux of 45 $\times 10^{9}$ molecules cm$^{-2}$ s$^{-1}$ ($\times$15 Earth’s global volcanic sulphur flux).}
\end{figure*}

Figure \ref{fig:Profiles_G_x20_F_x50} presents two profile plots from our HSO cross-section sensitivity tests, for the Sun (on the left), with SO$_2$ = 6 $\times 10^{10}$ molecules cm$^{-2}$ s$^{-1}$, and $\epsilon$ Eridani (on the right), SO$_2$ = 45 $\times 10^{9}$ molecules cm$^{-2}$ s$^{-1}$. The panels show the altitude-dependent mixing ratios of key atmospheric species computed using the default (dashed lines) and updated (solid lines) HSO cross-section prescriptions.

We conducted HSO cross-section sensitivity tests across the three FGK-type stellar scenarios, and did not identify differences in trace species abundances for a planet orbiting an F-type host, either when adopting the updated cross sections in place of the default prescription or when applying scaled versions of the updated cross sections. This behaviour reflects the sufficiently high UV–Visible flux provided by this stellar type, which drives HSO photolysis rates into a saturated regime where the associated photochemistry is not limited by the adopted cross-section magnitude or spectral coverage.

Overall, differences in trace gas abundance profiles modelled using the updated and default HSO cross-section prescriptions remain small across all atmospheric layers, staying well below one order of magnitude for all gaseous species, and for both G- and K-type stars. For CO$_2$, the impact of the updated prescription is negligible, with both profiles essentially overlapping. Even when we consider a scenario in which our new prescription underestimates real HSO cross sections by a factor of 10$^3$, or overestimates them by a factor of 10$^{-3}$ (i.e., where scaled prescriptions could more closely approach real values), the observed CO$_2$ profile remains unaffected. This reflects that the abundance of CO$_2$ is not significantly influenced by HSO photolysis.

For the remaining gaseous species, H$_2$O, CH$_4$, and CO, the updated HSO cross sections generally lead to a slight decrease in their mixing ratios. An exception occurs for H$_2$O in the G-type scenario, where its abundance increases near the tropopause, at altitudes of $\sim$11–20 km. As observed for CO$_2$, the abundance profiles of these species are weakly sensitive to substantial under- or overestimates of the HSO UV–visible cross sections in both the G- and K-type stellar scenarios. In practice, scaling the updated cross sections upward to account for a potential underestimate produces abundance profiles that are nearly indistinguishable from those obtained with the unscaled updated prescription, while scaling them downward to account for a potential overestimate yields profiles that closely resemble the default HO$_2$ case.

\begin{figure*}[ht!]
\centering
\includegraphics[width=\textwidth]{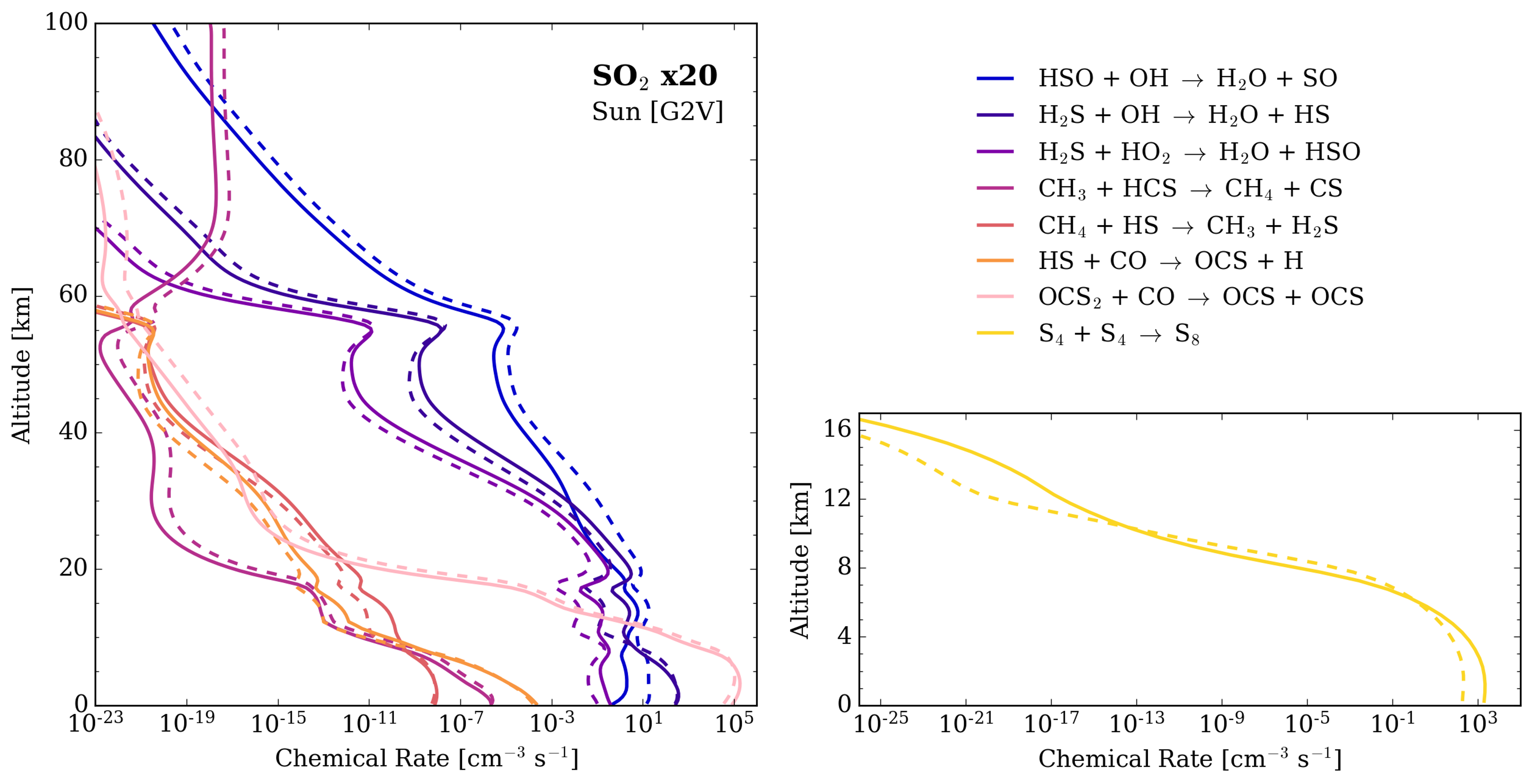}
\caption{Reaction rates for an anoxic, Archean-like planet orbiting the Sun, with an SO$_2$ surface flux of 6 $\times$ 10$^{10}$ molecules cm$^{-2}$ s$^{-1}$. Solid lines show reaction rates modelled using the updated HSO cross sections, while dashed lines use HO$_2$ cross sections. The plot on the left shows the reaction rates up to 100 km; on the right, the rates up to 17 km are shown.\label{fig:reaction_rates_for_key_reactions}}
\end{figure*}

Figure \ref{fig:reaction_rates_for_key_reactions} shows relevant HSO photolysis-sensitive chemical reaction rates for the same simulations as in Figure \ref{fig:Profiles_G_x20_F_x50} for the G-type host. Dashed and solid lines correspond to models using the default and updated HSO prescriptions, respectively. Differences in the modelled H$_2$O abundance profiles are largely influenced by a decrease in the reaction rate of HSO + OH. The enhanced HSO photolysis resulting from our updated cross sections promotes the depletion of HSO in the atmosphere, which reduces its availability for reaction with OH and the subsequent formation of H$_2$O. Moreover, we observe an increase in the reaction rates of H$_2$S + OH and H$_2$S + HO$_2$ near the tropopause, which contributes to the enhanced water abundance estimated at these altitudes. At the same time, the updated cross sections limit the accumulation of HCS in the atmosphere, which suppresses CH$_4$ production via its reaction with CH$_3$. Additionally, enhanced HSO photolysis increases HS production, which in turn favours CH$_4$ destruction via the CH$_4$ + HS reaction.

Alongside the effects on gaseous species, the updated HSO cross sections also influence aerosol chemistry, most notably the abundance of S$_8$. As briefly mentioned in Section \ref{sec:Intro}, HSO formation is connected to SO$_2$ photolysis via:
\begin{align*}
\text{SO}_2 + h\nu &\rightarrow \text{SO} + \text{O} \\
\text{SO} + \text{H} + \text{M} &\rightarrow \text{HSO} + \text{M} \\[-4pt]
&\hspace{-7em}\rule{13em}{0.4pt}\\
\text{\textbf{Net:}}\quad \text{SO}_2 + \text{H} &\rightarrow \text{HSO} + \text{O}
\end{align*}

Upon formation, HSO can either be recycled back to SO$_2$ (see Appendix \ref{appendixC}) or undergo photolysis. The latter produces HS, a key intermediate in the sulphur polymerisation network leading to S$_8$ formation. The overall S$_8$ formation pathway can be summarised as:
\begin{align*}
8 \times (\text{HSO} + h\nu &\rightarrow \text{HS} + \text{O}) \\
8 \times (\text{HS} + \text{H} &\rightarrow \text{H}_2 + \text{S}) \\
4 \times (\text{S} + \text{S} + \text{M} &\rightarrow \text{S}_2 + \text{M}) \\
2 \times (\text{S}_2 + \text{S}_2 + \text{M} &\rightarrow \text{S}_4 + \text{M}) \\
\text{S}_4 + \text{S}_4 + \text{M} &\rightarrow \text{S}_8 + \text{M} \\[-4pt]
&\hspace{-8.6em}\rule{17em}{0.4pt}\\
\textbf{Net:}\quad 8\,\text{HSO} + 8\,\text{H} &\rightarrow 8\,\text{H}_2 + \text{S}_8 + 8\,\text{O}
\end{align*}

Importantly, the higher HSO photolysis rates resulting from the updated cross sections enhance HS production, ultimately yielding S$_8$ abundances up to four orders of magnitude larger than those obtained with the default prescription (see Figure \ref{fig:Profiles_G_x20_F_x50}, lower panels). At the same time, this shifts the sulphur budget away from pathways that recycle HSO back into SO$_2$ or oxidise SO$_2$ into H$_2$SO$_4$ aerosols (see Appendix \ref{appendixC}), therefore reducing the efficiency of alternative channels for atmospheric sulphur removal via subsequent rainout or sedimentation of these species.

Notably, the K-type host exhibits larger differences in tropospheric S$_8$ abundances between the default and updated cross-section prescriptions than those seen for the G-type case. This reflects the weaker UV flux of the K-type star, under which the default HO$_2$-based prescription captures only a limited fraction of the HSO photolysis potential. Extending the HSO cross sections into the 250–465 nm range, where $\epsilon$ Eridani still provides appreciable flux and SO$_2$ opacity is low, represents a proportionally larger enhancement to the photolysis rate than in the G-type scenario.

\subsection{Spectral Sensitivity}

\begin{figure*}[t]
    \centering
    \includegraphics[width=0.9\textwidth]{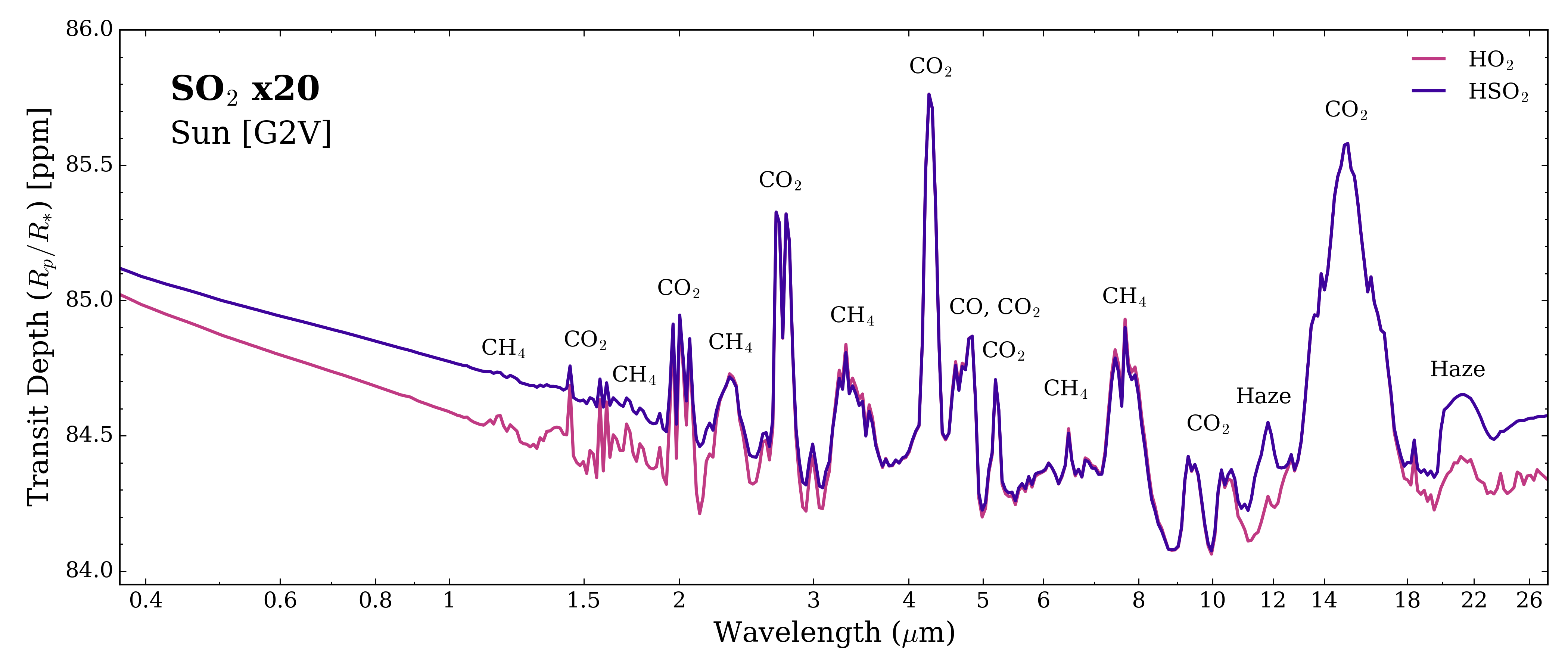}
    \includegraphics[width=0.9\textwidth]{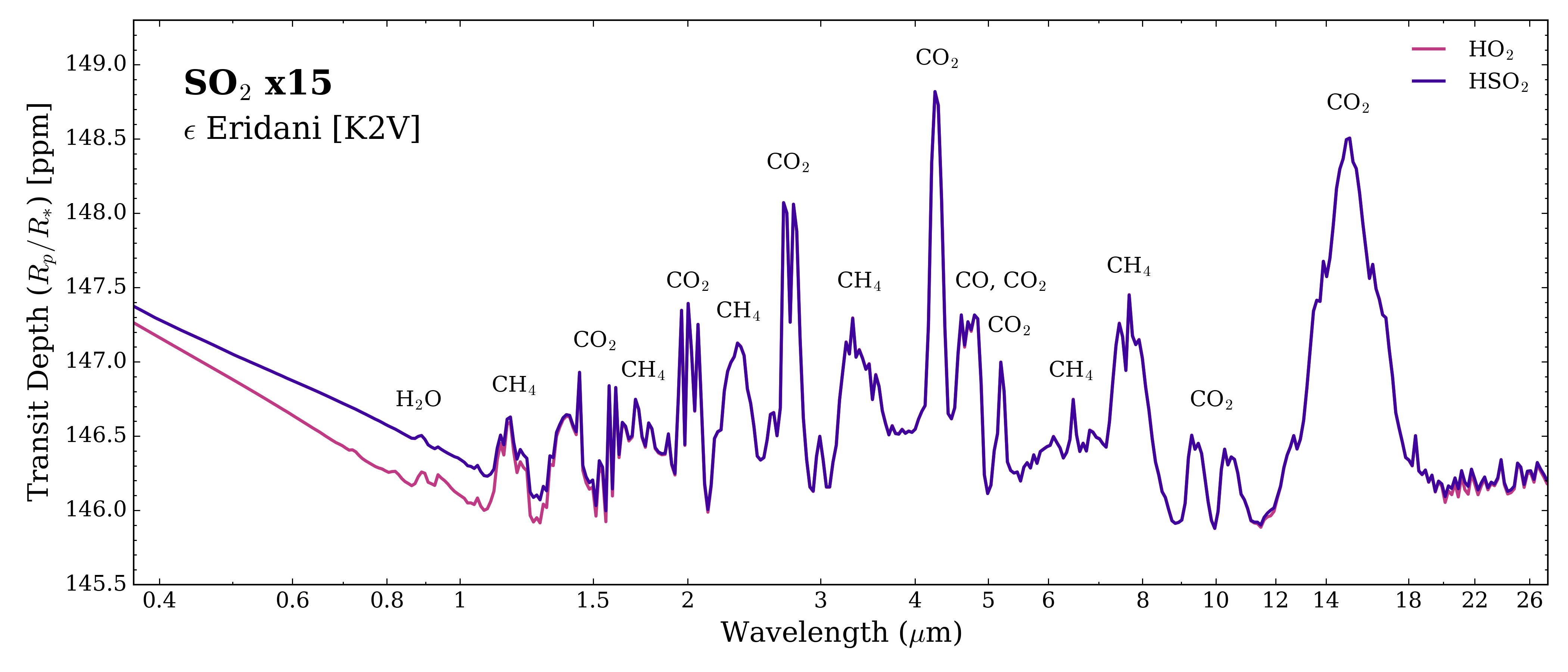}
    \caption{Top: Comparison of the transmission spectra modelled using the default (HO$_2$) and updated (HSO$_2$) HSO cross-section prescriptions for a planet orbiting the Sun with a surface SO$_2$ flux of $\times 20$ Earth’s global volcanic sulphur flux (SO$_2$ = 6 $\times 10^{10}$ molecules cm$^{-2}$ s$^{-1}$). Bottom: Same as the plot in the upper panel, for a planet orbiting $\epsilon$ Eridani, with an SO$_2$ surface flux of $\times 15$ Earth’s global volcanic sulphur flux (SO$_2$ = 45 $\times 10^{9}$ molecules cm$^{-2}$ s$^{-1}$).}
    \label{fig:transmission_g_f}
\end{figure*}

\begin{figure*}[t]
    \centering
    \includegraphics[width=0.9\textwidth]{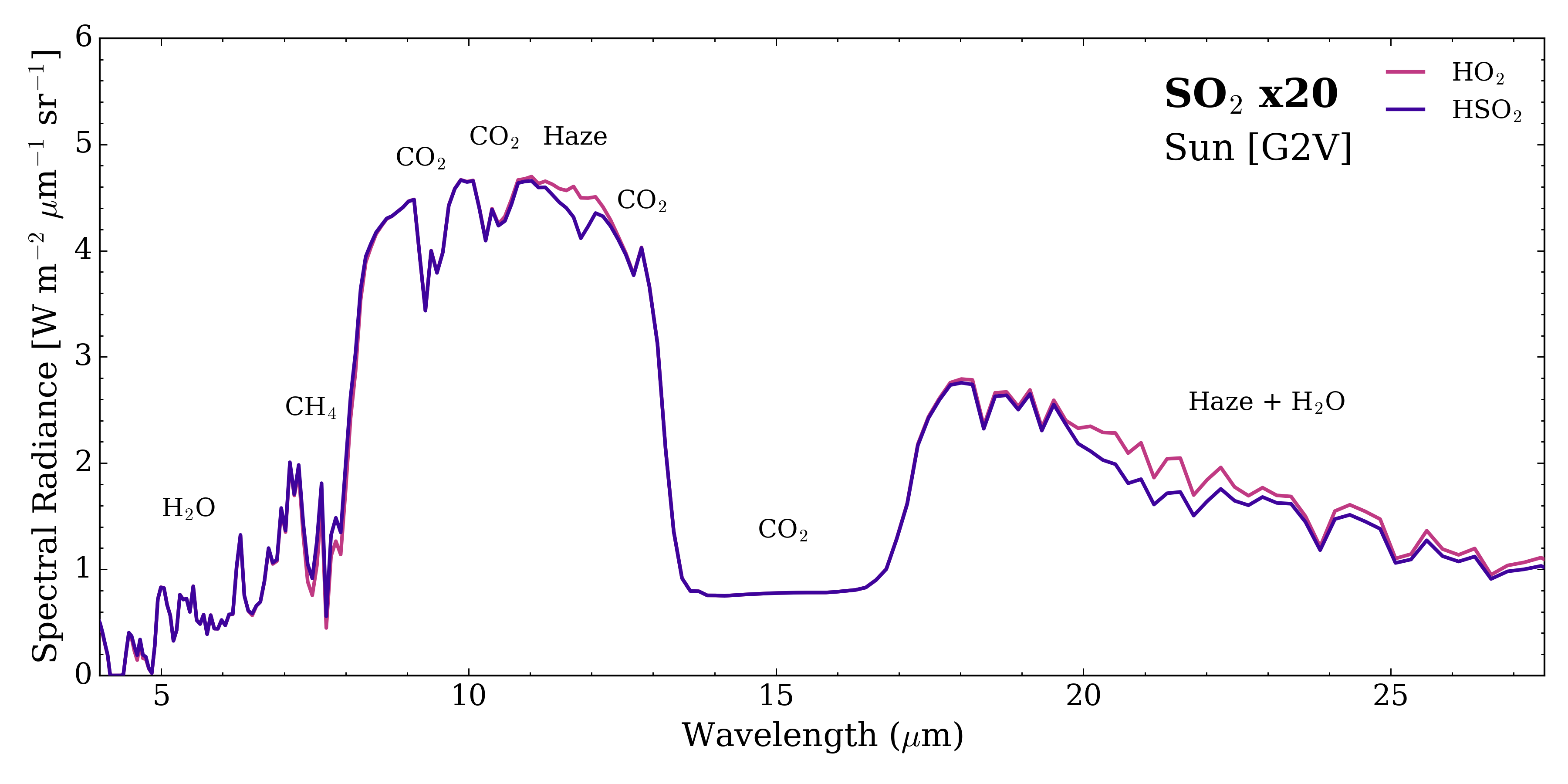}
    \includegraphics[width=0.9\textwidth]{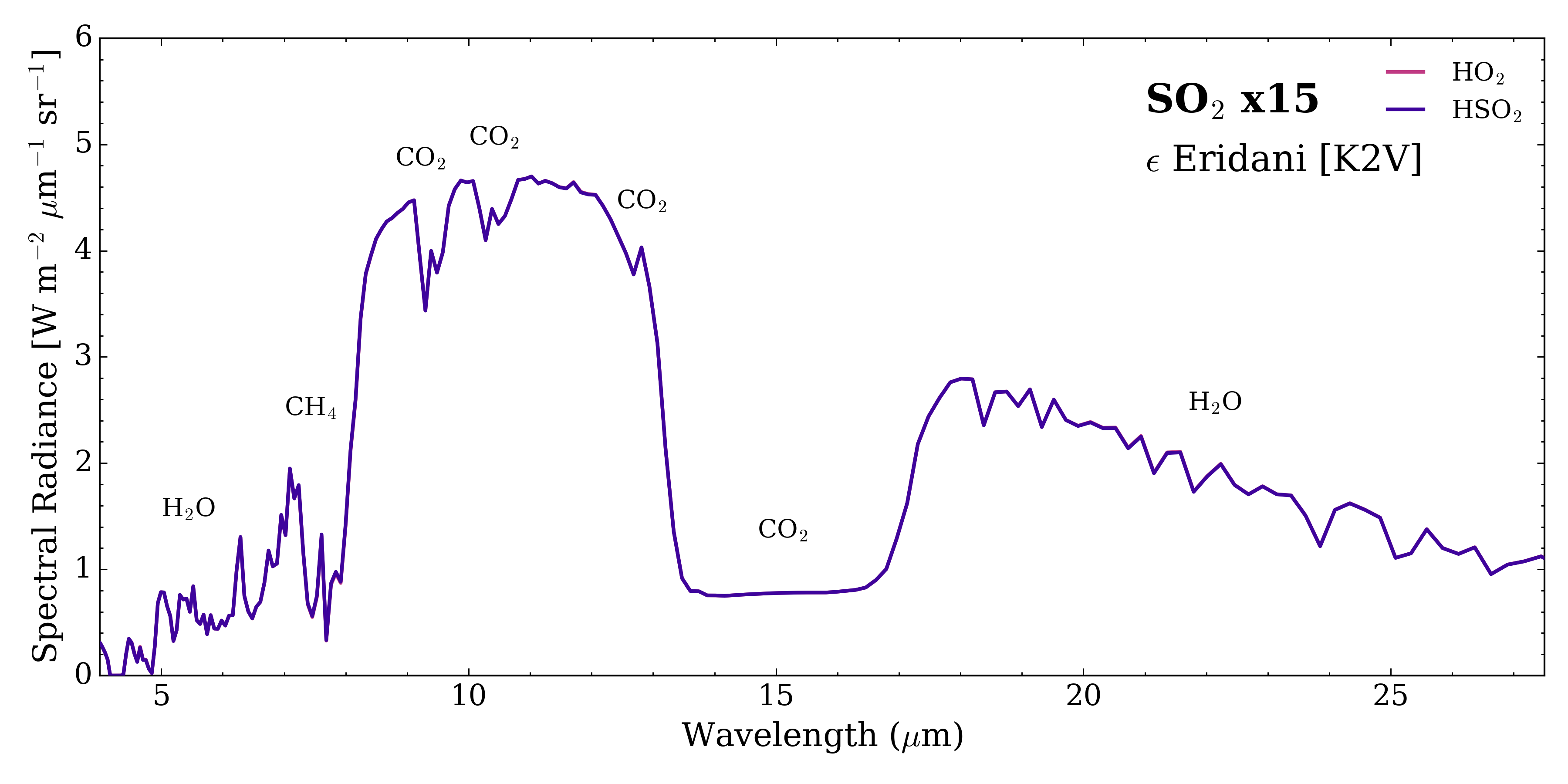}
    \caption{Top: Comparison of the emission spectra modelled using the default (HO$_2$) and updated (HSO$_2$) HSO cross-section prescriptions for a planet orbiting the Sun with a surface SO$_2$ flux of $\times 20$ Earth’s global volcanic sulphur flux (SO$_2$ = 6 $\times 10^{10}$ molecules cm$^{-2}$ s$^{-1}$). Bottom: Same as the plot in the upper panel, for a planet orbiting $\epsilon$ Eridani, with an SO$_2$ surface flux of $\times 15$ Earth’s global volcanic sulphur flux (SO$_2$ = 45 $\times 10^{9}$ molecules cm$^{-2}$ s$^{-1}$).}
    \label{fig:emission_g_f}
\end{figure*}

\begin{figure*}[t]
    \centering
    \includegraphics[width=0.9\textwidth]{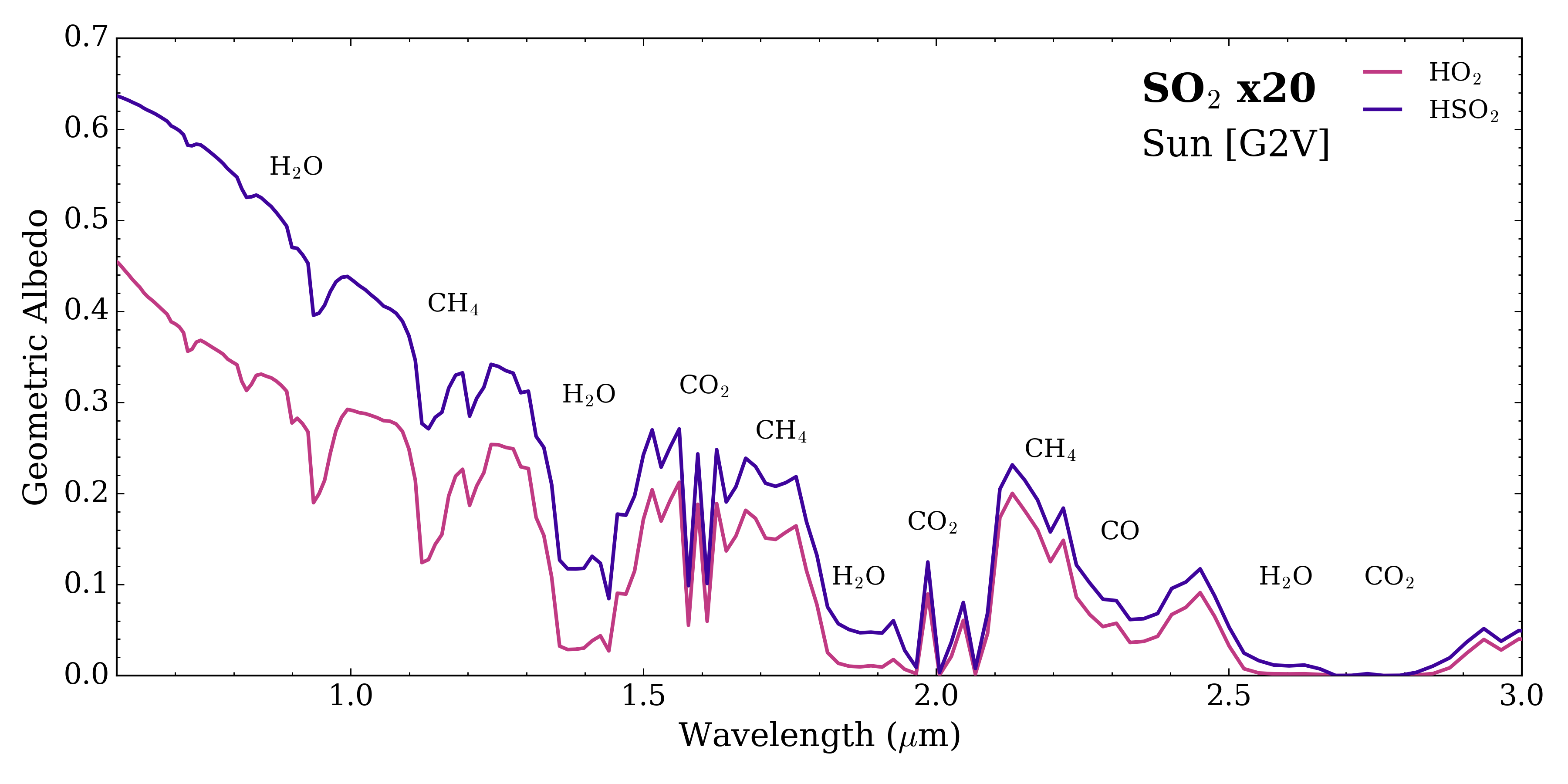}
    \includegraphics[width=0.9\textwidth]{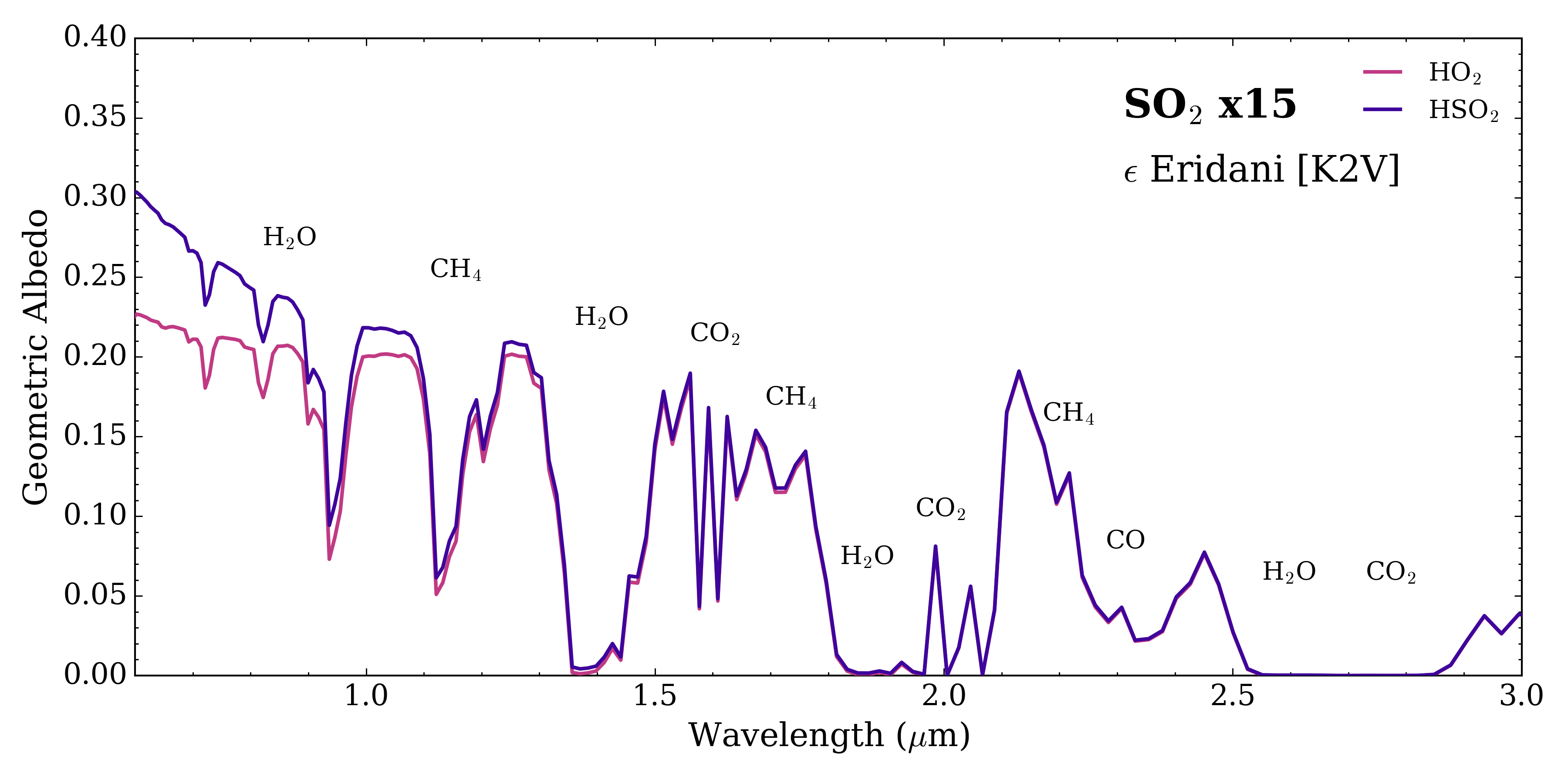}
    \caption{Top: Comparison of the reflection spectra modelled using the default (HO$_2$) and updated (HSO$_2$) HSO cross-section prescriptions for a planet orbiting the Sun with a surface SO$_2$ flux of $\times 20$ Earth’s global volcanic sulphur flux (SO$_2$ = 6 $\times 10^{10}$ molecules cm$^{-2}$ s$^{-1}$). Bottom: Same as the plot in the upper panel, for a planet orbiting $\epsilon$ Eridani, with an SO$_2$ surface flux of $\times 15$ Earth’s global volcanic sulphur flux (SO$_2$ = 45 $\times 10^{9}$ molecules cm$^{-2}$ s$^{-1}$).}
    \label{fig:reflection_g_f}
\end{figure*}

Differences in the abundance profiles of trace-gas species, driven by the choice of a given HSO cross-section prescription, can produce differences in the resulting transmission, emission, and reflection spectra of terrestrial, Archean-like planets. Figures \ref{fig:transmission_g_f}--\ref{fig:reflection_g_f} show simulated transmission, emission, and reflection spectra for planets orbiting a G-type and a K-type host, using the default and updated HSO cross sections. We have labelled spectral features in all forward models, identifying those produced by gaseous species as well as those linked to S$_8$ aerosols, which we denote as ``Haze''.

In the transmission spectra, we find that the updated HSO cross sections produce more pronounced scattering slopes at optical and near-infrared wavelengths, along with stronger far-infrared absorption features associated with S$_8$ aerosols (at around 12 $\mu$m and 21 $\mu$m). These differences reflect the higher S$_8$ abundances produced with the updated cross sections compared to the default prescription, and are more evident for a planet around a G-type star, whose higher UV-Visible flux drives more efficient HSO photolysis and S$_8$ formation. This sensitivity is also further increased by the larger SO$_2$ surface flux used in the G-type scenario.

Similarly, in the emission spectra, the updated HSO cross-section prescription can produce sufficiently high haze opacities for the S$_8$ absorption features at $\sim 12$ $\mu$m and $\sim 21$ $\mu$m to become visible in the G-type host scenario. The same enhancement in S$_8$ abundance under the new prescription is responsible for the higher geometric albedos in the reflection spectra, for both stellar types.

\section{Discussion} \label{sec:Discussion}

The impact of our updated HSO cross sections is consequential for anoxic, Archean Earth-like planets orbiting both G- and K-type stars. In addition to the FGK-type scenarios, we note that we also ran sensitivity tests for an M-type host, which revealed no differences between both cross-section prescriptions independently of the SO$_2$ surface flux.

The absence of any changes in the M-type scenario follows naturally from the host star's markedly lower UV–Visible flux relative to Solar-type stars (see Figure \ref{fig:stellar_flux_hso_xsec}), resulting in negligible HSO photolysis rates, and leaving the atmosphere insensitive to which cross-section prescription is used. Given that rocky planets orbiting M-dwarfs are prime targets for atmospheric characterization with JWST, this is a promising outcome of our study, as it suggests that the interpretation of their spectra should be robust to the choice of HSO cross-section prescriptions.

The characterization of temperate, rocky exoplanets around Solar-type stars, on the other hand, will be the focus of future missions such as the Habitable Worlds Observatory (HWO) and the Large Interferometer For Exoplanets (LIFE) \citep{Quanz2022,DecadalSurveyHWO}. HWO will be optimized for reflected light observations, with potential for transmission spectroscopy studies depending on the final telescope design \citep{Mamajek_Stapelfeldt2024}. With an expected spectral coverage spanning UV, optical, and infrared wavelengths, the interpretation of HWO observations is susceptible to variations in HSO cross sections. Conversely, LIFE will target terrestrial planets in emission. In this case, our results suggest that the choice of HSO cross-section prescription may influence the detectability of S$_8$ aerosol features at $\sim 12$ $\mu$m and $\sim 21$ $\mu$m in observations of exoplanets orbiting G-type host stars (see Figure \ref{fig:emission_g_f}).

Importantly, our results show that the updated HSO prescription results in higher atmospheric abundances of S$_8$ aerosols. Because S$_8$ is highly resistant to photodissociation, it is a key tracer for inferring volcanic sulphur emissions on anoxic terrestrial exoplanets, effectively acting as an indirect signature of ongoing volcanic outgassing. This same UV resilience allows S$_8$ aerosols to provide substantial shielding of the planetary surface from incident ultraviolet radiation. Such shielding may influence the surface UV environment relevant for prebiotic chemistry, particularly in shallow water settings, by attenuating wavelengths that drive or inhibit specific chemical pathways.

Furthermore, variations in S$_8$ aerosol abundances resulting from distinct HSO cross-section prescriptions have direct consequences for the deposition rates of this species onto planetary surfaces, with implications for signatures of sulphur mass-independent fractionation (S-MIF). For the planetary scenario orbiting a G-type host star, as an example, the default HSO prescription based on HO$_2$ data yields an S$_8$ deposition rate one order of magnitude lower than that obtained using our updated prescription. Signatures of S-MIF are preserved in the geological record of the Archean Earth and have been tentatively detected in Martian meteorites, providing key constraints on atmospheric chemistry and climate evolution of both planets \citep{PavlovKasting2002,Franz2014,Tomkins2020,CatlingZahnle2020}.

Previous photochemical studies of volcanically driven anoxic atmospheres on ancient Mars predicted the formation and deposition of S$_8$ under such conditions, and proposed it as a potential mineralogical signature of reducing environments \citep{Sholes2017}. The subsequent \textit{in situ} detection of elemental sulphur crystals made by the Curiosity rover demonstrates that S$_8$ could indeed accumulate and be preserved at the Martian surface \citep{Sklute2024AGU}. Although the specific geochemical features detected by Curiosity may favour hydrothermal pathways rather than a pure atmospheric chemistry origin, this observation underscores the importance of constraining the relative contributions of atmospheric and geological processes in shaping the Martian sulphur inventory and potential S-MIF signatures.

Moreover, photochemical models have persistently struggled to reproduce key aspects of the Archean S-MIF record \citep[e.g.,][]{Harman2018}, potentially highlighting model sensitivity to a relatively large range of atmospheric parameters, including poorly constrained atmospheric reaction rates. Our results suggest that HSO cross-section data may have a non-negligible impact on S-MIF, and motivates future work using updated sulphur photochemistry.

As in previous studies \citep{Ranjan2020,Broussard2024,Broussard2025}, we have only considered the impacts of updated HSO cross-section prescriptions on the trace-gas abundances of Archean Earth-like, N$_2$–CO$_2$–H$_2$O–dominated atmospheres. This represents only one of the many planetary scenarios that may exist among the diverse population of exoplanets. Different compositions will be more or less affected by changes in the HSO cross sections. Indeed, even within planetary archetypes sharing this composition, distinct sulphur surface fluxes can significantly influence the sensitivity of the atmosphere to HSO photochemistry. As an example, planets with substantially lower SO$_2$ emissions than modern Earth would be essentially insensitive to variations in the HSO cross sections, as HSO production in such atmospheres would occur at much smaller scales than those considered here.

It is also important to note that the updated HSO cross-section prescription presented in this study is a proxy, derived from HSO$_2$ data, and does not represent direct measurements of HSO itself. Any proxy data will inevitably deviate from actual UV–Visible cross sections of HSO, even when the choice of a proxy molecule is carefully justified. Overall, improving the accuracy of forward models and atmospheric retrieval codes across diverse planetary environments will ultimately require reliable UV–visible cross-section data for all molecules and associated intermediate species relevant to the atmospheres modelled.

An additional limitation of this study is that the cross sections we use to model HSO photolysis are not temperature dependent. HO$_2$ cross sections have been measured at 298 K \citep{Sander2011}, while HSO$_2$ data has been estimated through ab initio calculations also performed at a single temperature \citep{Lu2021}. The temperature dependence of HSO cross sections may prove particularly important for the characterization of closer-in rocky planets (e.g., hot super-Earths), where enhanced sulphur outgassing could amplify the impact of HSO photochemistry. Ideally, temperature-dependent cross sections should be directly measured or computed and incorporated in atmospheric models.

This analysis has not looked into the sensitivity of HSO cross-section prescriptions to chemical reaction rates. These also constitute fundamental inputs in photochemical models and have been identified as requiring further investigation for certain reactions \citep[e.g.,][]{Ranjan2020}, which may further affect atmospheric abundances. Similarly, the effects of collisional broadening coefficients in HSO opacities arising from the pressure of dominant background gases were not considered in this study. Accounting for these effects would require temperature-dependent laboratory measurements or theoretical calculations of HSO broadening coefficients for relevant atmospheric compositions (e.g., an N$_2$-dominated background for an Archean-like environment), which are non-trivial to perform \citep{Niraula2022,Wiesenfeld2025}. Nevertheless, such effects may further influence the estimated abundances of trace species, although they are expected to have a fairly minor impact.

\section{Conclusions} \label{sec:Conclusions}

This work defines HSO$_2$ as a physically motivated proxy for HSO, improving upon the use of HO$_2$ UV-Visible cross sections for modelling its photolysis. By delineating the wavelength range over which HSO is expected to be spectrally active, we help constrain future experimental and {\it{ab initio}} studies aimed at characterizing its UV-Visible absorption spectrum.

Additionally, this analysis has tested the impact of different HSO cross-section prescriptions on the abundances of trace species in anoxic, Archean-like atmospheres of planets orbiting Sun-like stars (FGK-type). Specifically, we compared data from the default HSO proxy, HO$_2$ \citep{Sander2011}, with that of a newly proposed proxy, HSO$_2$ \citep{Lu2021}. We focused on scenarios with elevated SO$_2$ surface fluxes, and on Solar-type hosts in order to maximize model sensitivity to variations in HSO cross sections and identify conditions under which these variations could produce the most significant changes in atmospheric composition and potential observational signatures.

We find that differences in abundance profiles calculated using the two prescriptions remain well below an order of magnitude for most atmospheric species across the simulated stellar types. The exception is S$_8$ aerosols in the troposphere of planets orbiting G- and a K-type stars, which can increase in abundance by up to four orders of magnitude compared to predictions using the default prescription. Even when accounting for a potential underestimation of values in our updated HSO cross sections by a factor of 10$^3$, the resulting trace species abundances remain essentially indistinguishable from those obtained with the unscaled updated prescription.

We report changes in spectral observables in simulated transmission, emission, and reflection spectra that are generally linked to the presence of S$_8$ atmospheric hazes and to the sensitivity of their abundances to the adopted HSO prescription. Depending on the instrumental sensitivity of upcoming spectroscopic facilities, these variations may translate into observable differences in the resulting planetary spectra.

Overall, this work underscores the importance of sensitivity tests in the prioritization of molecular species for detailed spectral characterization, particularly in the UV–Visible range. Ultimately, molecular absorption cross sections represent critical inputs in atmospheric modelling, and their accuracy is essential for a reliable interpretation of exoplanet spectra.

\appendix

\section{Biotic Atmospheric Boundary Conditions} \label{appendixA}

Table \ref{table:boundary_conditions} lists the biotic atmospheric boundary conditions used for modelling results from the main text of this paper. These are the surface fluxes, surface volume mixing ratios, and dry deposition velocities where relevant.

\begin{table}[h]
\centering
\fontsize{9pt}{10.5pt}\selectfont
\caption{Biotic atmospheric species boundary conditions, including prescribed surface fluxes, surface volume mixing ratios, and dry deposition velocities. \label{table:boundary_conditions}}
\scriptsize
\renewcommand{\arraystretch}{0.84}
\begin{tabularx}{\textwidth}{l *{7}{>{\centering\arraybackslash}X}}
\toprule
\textbf{Species} & \makecell{\textbf{Surface Flux}\\[0.em]\textbf{[molecules cm$^{-2}$ s$^{-1}$]}} & \makecell{\textbf{Surface Mixing}\\[0.em]\textbf{Ratio [v/v]}} & \makecell{\textbf{Dry Deposition}\\[0.em]\textbf{Velocity [cm s$^{-1}$]}}\\
\midrule
\textbf{O} & 0 & \textbf{--} & 1\\
\textbf{O$_2$} & 0 & \textbf{--} & 0\\
\textbf{H$_2$O} & 0 & \textbf{--} & 0\\
\textbf{H} & 0 & \textbf{--} & 1\\
\textbf{OH} & 0 & \textbf{--} & 1\\
\textbf{HO$_2$} & 0 & \textbf{--} & 1\\
\textbf{H$_2$O$_2$} & 0 & \textbf{--} & 2$\times$10$^{-1}$\\
\textbf{H$_2$} & 1$\times$10$^{10}$ & \textbf{--} & 2.4$\times$10$^{-4}$\\
\textbf{CO} & 1$\times$10$^{8}$ & \textbf{--} & 1.2$\times$10$^{-4}$\\
\textbf{HCO} & 0 & \textbf{--} & 1\\
\textbf{H$_2$CO} & 0 & \textbf{--} & 2$\times$10$^{-1}$\\
\textbf{CH$_4$} & 1.1$\times$10$^{10}$ & \textbf{--} & 0\\
\textbf{CH$_3$} & 0 & \textbf{--} & 1\\
\textbf{C$_2$H$_6$} & 0 & \textbf{--} & 0\\
\textbf{NO} & 0 & \textbf{--} & 3$\times$10$^{-4}$\\
\textbf{NO$_2$} & 0 & \textbf{--} & 3$\times$10$^{-3}$\\
\textbf{HNO} & 0 & \textbf{--} & 1\\
\textbf{O$_3$} & 0 & \textbf{--} & 7$\times$10$^{-2}$\\
\textbf{HNO$_3$} & 0 & \textbf{--} & 2$\times$10$^{-1}$\\
\textbf{N} & 0 & \textbf{--} & 0\\
\textbf{C$_3$H$_2$} & 0 & \textbf{--} & 0\\
\textbf{C$_3$H$_3$} & 0 & \textbf{--} & 0\\
\textbf{CH$_3$C$_2$H} & 0 & \textbf{--} & 0\\
\textbf{CH$_2$CCH$_2$} & 0 & \textbf{--} & 0\\
\textbf{C$_3$H$_5$} & 0 & \textbf{--} & 0\\
\textbf{C$_3$H$_6$} & 0 & \textbf{--} & 0\\
\textbf{C$_3$H$_7$} & 0 & \textbf{--} & 0\\
\textbf{C$_3$H$_8$} & 0 & \textbf{--} & 0\\
\textbf{C$_2$H$_4$OH} & 0 & \textbf{--} & 0\\
\textbf{C$_2$H$_2$OH} & 0 & \textbf{--} & 0\\
\textbf{C$_2$H$_5$} & 0 & \textbf{--} & 0\\
\textbf{C$_2$H$_4$} & 0 & \textbf{--} & 0\\
\textbf{CH} & 0 & \textbf{--} & 0\\
\textbf{CH$_3$O$_2$} & 0 & \textbf{--} & 0\\
\textbf{CH$_3$O} & 0 & \textbf{--} & 0\\
\textbf{CH$_2$CO} & 0 & \textbf{--} & 0\\
\textbf{CH$_3$CO} & 0 & \textbf{--} & 0\\
\textbf{CH$_3$CHO} & 0 & \textbf{--} & 0\\
\textbf{C$_2$H$_2$} & 0 & \textbf{--} & 0\\
\textbf{CH$_2$3} & 0 & \textbf{--} & 0\\
\textbf{C$_2$H} & 0 & \textbf{--} & 0\\
\textbf{C$_2$} & 0 & \textbf{--} & 0\\
\textbf{C$_2$H$_3$} & 0 & \textbf{--} & 0\\
\textbf{HCS} & 0 & \textbf{--} & 0\\
\textbf{CS$_2$} & 0 & \textbf{--} & 0\\
\textbf{CS} & 0 & \textbf{--} & 0\\
\textbf{OCS} & 0 & \textbf{--} & 0\\
\textbf{S} & 0 & \textbf{--} & 0\\
\textbf{HS} & 0 & \textbf{--} & 0\\
\textbf{H$_2$S} & 3.5$\times$10$^{8}$ & \textbf{--} & 2$\times$10$^{-2}$\\
\textbf{SO$_3$} & 0 & \textbf{--} & 0\\
\textbf{HSO} & 0 & \textbf{--} & 1\\
\textbf{H$_2$SO$_4$} & 0 & \textbf{--} & 1\\
\textbf{SO$_2$} & 3$\times$10$^{9}$ -- 3$\times$10$^{12}$ & \textbf{--} & 1\\
\textbf{SO} & 0 & \textbf{--} & 0\\
\textbf{CO$_2$} & 0 & 2$\times$10$^{-2}$ & 0\\
\textbf{SO$_4$ AER} & 0 & \textbf{--} & 10$^{-2}$\\
\textbf{S$_8$ AER} & 0 & \textbf{--} & 10$^{-2}$\\
\textbf{HC AER} & 0 & \textbf{--} & 10$^{-2}$\\
\textbf{HC AER2} & 0 & \textbf{--} & 10$^{-2}$\\
\bottomrule
\end{tabularx}
\end{table}

\section{HSO Photolysis Sensitivity to SO$_2$ Surface Flux and Cross-Section Scaling} \label{appendixB}

\begin{figure}[ht!]
\centering
\includegraphics[width=0.47\textwidth]{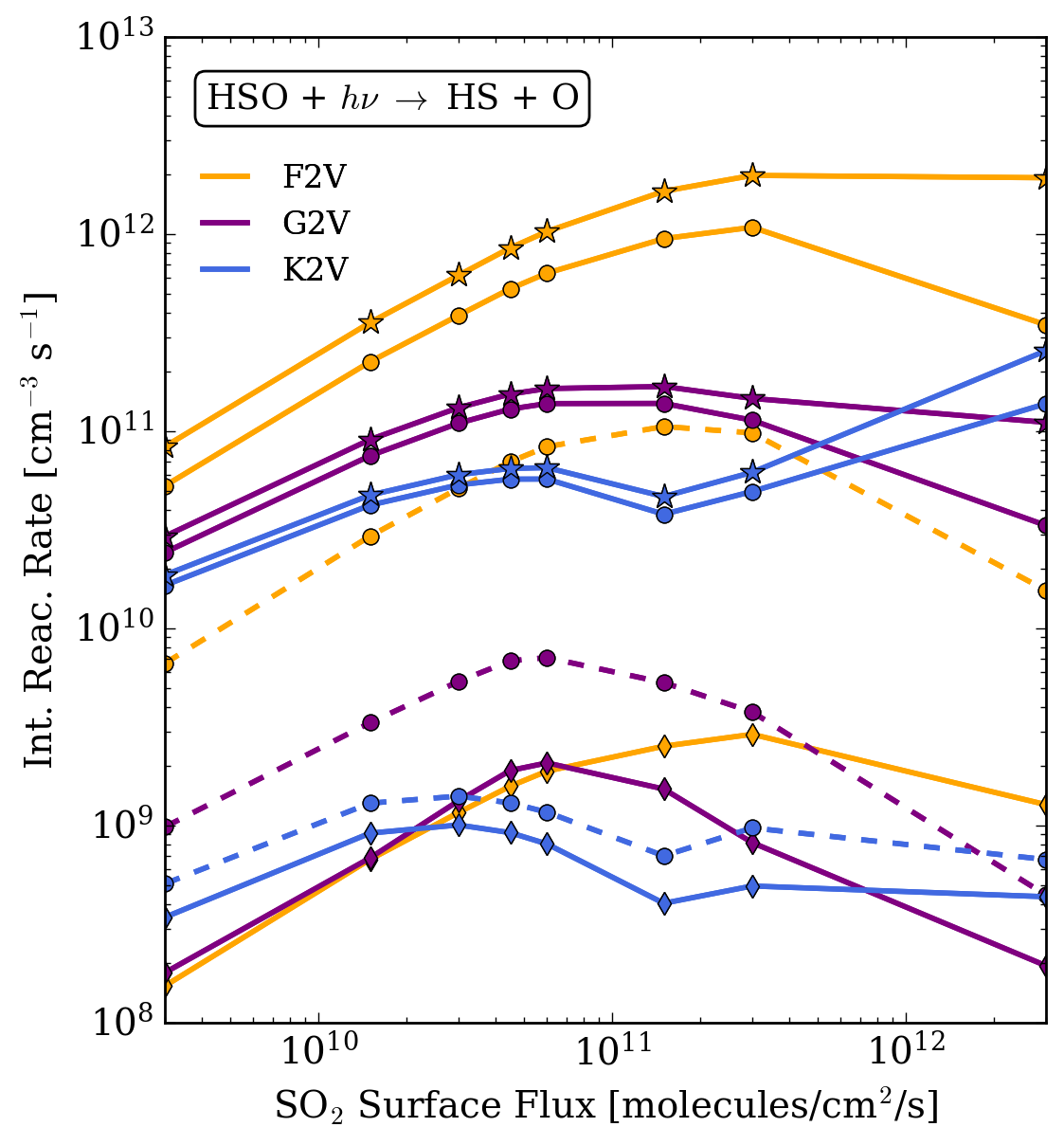}
\caption{Integrated HSO photolysis reaction rates as a function of SO$_2$ surface fluxes, ranging from $\times$1 to $\times$10$^3$ Earth’s global volcanic sulphur flux, for an anoxic planet orbiting $\sigma$ Böotis (F2V, in yellow), the Sun (G2V, in purple) and $\epsilon$ Eridani (K2V, in blue). Solid lines with circular markers show rates modelled using the updated HSO cross sections, while dashed lines with circular markers use HO$_2$ data. Solid lines with star and diamond markers correspond to scaled versions of the updated prescription, obtained by multiplying the cross sections by factors of 10$^{3}$ and 10$^{-3}$, respectively.} \label{fig:Int_HSO_PhotRate_FGK_appendix}
\end{figure}

Figure \ref{fig:Int_HSO_PhotRate_FGK_appendix} extends Figure \ref{fig:Int_HSO_PhotRate_FG} by exploring the dependence of integrated HSO photolysis reaction rates on scaled versions of the updated HSO cross-section prescription, for SO$_2$ surface fluxes ranging from $\times$1 to $\times$10$^{3}$ Earth’s global volcanic sulphur flux. In addition to the default and updated prescriptions, it includes scaled cross sections obtained by multiplying the updated values by factors of $10^{3}$ and $10^{-3}$.

\section{Alternative Sulphur Pathways Competing with S$_8$ Formation} \label{appendixC}

There are three alternative pathways through which sulphur injected into the atmosphere via SO$_2$ emissions can be removed without ultimately contributing to the formation and surface deposition of S$_8$ aerosols. Firstly, HSO can participate in chemical reactions that recycle sulphur back into SO$_2$, for example:
\begin{equation*}
\ce{HSO + OH -> H_2O + SO}
\end{equation*}
\vspace{-13pt}
\begin{equation*}
\ce{HSO + H -> H_2 + SO}
\end{equation*}

The resulting SO can subsequently be converted back into SO$_2$ via reactions such as:
\begin{equation*}
\ce{SO + OH -> SO_2 + H}
\end{equation*}
\vspace{-13pt}
\begin{equation*}
\ce{SO + SO -> SO_2 + S}
\end{equation*}

In addition, SO$_2$ can be oxidised towards the formation of H$_2$SO$_4$ aerosols through chemical pathways including:
\begin{equation*}
\ce{SO_2 + OH + M -> HSO_3 + M}
\end{equation*}
\vspace{-13pt}
\begin{equation*}
\ce{HSO_3 + H -> H_2 + SO_3}
\end{equation*}
\vspace{-13pt}
\begin{equation*}
\ce{SO_3 + H_2O + M -> H_2SO_4 + M}
\end{equation*}

SO$_2$ can also be regenerated through reactions involving SO$_3$, for example:
\begin{equation*}
\ce{SO_3 + $h\nu$ -> SO_2 + O}
\end{equation*}
\vspace{-13pt}
\begin{equation*}
\ce{SO_3 + SO -> SO_2 + SO_2}
\end{equation*}
\vspace{-13pt}
\begin{equation*}
\ce{SO_3 + CO -> SO_2 + CO_2}
\end{equation*}


\begin{acknowledgments}
This work was performed by the Experimental Constraints for Improving Terrestrial Exoplanet Photochemical Models (ExCITE-PM) Team funded by NASA Exoplanet Research Program (XRP) grant No. 80NSSC22K0235. A.B would like to acknowledge support from the Heising-Simons Foundation and Bard College through the co-funded fellowship 2023-4047. C.S.S. acknowledges the support from Fundação para a Ciência e Tecnologia (FCT) in the form of a work contract through the Scientific Employment Incentive program (DOI: 10.54499/2022.06872.CEECIND/CP1721/CT0002). O.D.S.D. acknowledges support from e-CHEOPS (PEA No 4000142255). This work was also supported by FCT - Fundação para a Ciência e a Tecnologia through national funds under grants No. UIDB/04434/2020 (DOI: 10.54499/UIDB/04434/2020), UIDP/04434/2020 (DOI: 10.54499/UIDP/04434/2020), and UID/04434/2025. The authors would also like to thank Christopher L. Strand and Frances M. Gomez for their  support and contribution.\\
\textit{Software}: Atmos \citep{Kasting1979,Arney2016}, POSEIDON \citep{MacDonald2017,MacDonald2023}.
\end{acknowledgments}

\bibliography{sample701}{}
\bibliographystyle{aasjournalv7}

\end{document}